\documentclass[  superscriptaddress, amsmath,amssymb, aps,longbibliography]{revtex4-1}
\usepackage{graphicx}
\usepackage{dcolumn}
\usepackage{amsthm}
\usepackage{bm}

\pdfoutput=1
\usepackage{graphicx,graphics,epsfig,subfigure,times,bm,bbm,amssymb,amsmath,amsthm,mathrsfs,MnSymbol}
\usepackage{gensymb}
\usepackage{amsfonts}
\usepackage[matrix,frame,arrow]{xypic}
\usepackage[pdfstartview=FitH]{hyperref}
\hypersetup{
    colorlinks=true,       
    linkcolor=red,          
    citecolor=magenta,        
    filecolor=magenta,      
    urlcolor=cyan,           
    runcolor=cyan}
\usepackage{epstopdf}
\usepackage[pdftex]{color}
\usepackage{braket}
\usepackage{enumerate}
\usepackage[normalem]{ulem}
\usepackage[usenames,dvipsnames]{xcolor}
\usepackage{multirow}
\usepackage{mathtools}

\definecolor{orange}{rgb}{1,0.5,0}

\newcommand{\ignore}[1]{}
\usepackage{geometry}

\ignore{
\documentclass[eprintnumbers,amsmath,amssymb,onecolumn,a4paper,caption ]{article}
\usepackage{amsfonts}
\usepackage{amssymb}
\usepackage{mathrsfs}
\usepackage{mathbbold}
\usepackage{bbm}
\usepackage{mathrsfs}
\usepackage{dcolumn}
\usepackage{bm}
\usepackage{times,epsfig,amssymb,amsmath}
\usepackage{float}
\usepackage{subfigure}
\usepackage{geometry}\geometry{left=2.5cm,right=2.5cm,top=3cm,bottom=3cm}

\usepackage{color}

}

\begin{document}

\title{Improving the performance of quantum approximate optimization for preparing non-trivial quantum states without translational symmetry}

\author{Zheng-Hang Sun}
\affiliation{Institute of Physics, Chinese Academy of Sciences, Beijing 100190, China}
\affiliation{School of Physical Sciences, University of Chinese Academy of Sciences, Beijing 100190, China}

\author{Yong-Yi Wang}
\email{yywang@iphy.ac.cn}
\affiliation{Institute of Physics, Chinese Academy of Sciences, Beijing 100190, China}
\affiliation{School of Physical Sciences, University of Chinese Academy of Sciences, Beijing 100190, China}

\author{Jian Cui}
\email{jiancui@buaa.edu.cn}
\affiliation{School of Physics, Beihang University, Beijing 100191, China}

\author{Heng Fan}
\email{hfan@iphy.ac.cn}
\affiliation{Institute of Physics, Chinese Academy of Sciences, Beijing 100190, China}
\affiliation{School of Physical Sciences, University of Chinese Academy of Sciences, Beijing 100190, China}
\affiliation{Songshan Lake  Materials Laboratory, Dongguan 523808, Guangdong, China}
\affiliation{CAS Center of Excellence for Topological Quantum Computation, University of Chinese Academy of Sciences,
Beijing 100190, China}
\affiliation{Beijing Academy of Quantum Information Sciences, Beijing 100193, China}

\begin{abstract}
\noindent The variational preparation of complex quantum states using the quantum approximate optimization algorithm (QAOA) is of fundamental interest, and becomes a promising application of quantum computers. Here, we systematically study the performance of QAOA for preparing ground states of target Hamiltonians near the critical points of their quantum phase transitions, and generating Greenberger-Horne-Zeilinger (GHZ) states. We reveal that the performance of QAOA is related to the translational invariance of the target Hamiltonian: Without the translational symmetry, for instance due to the open boundary condition (OBC) or randomness in the system, the QAOA becomes less efficient.  We then propose a generalized QAOA assisted by the parameterized resource Hamiltonian (PRH-QAOA), to achieve a better performance. In addition, based on the PRH-QAOA, we design a low-depth quantum circuit beyond one-dimensional geometry, to generate GHZ states with perfect fidelity. The experimental realization of the proposed scheme for generating GHZ states on Rydberg-dressed atoms is discussed. Our work paves the way for performing QAOA on programmable quantum processors without translational symmetry, especially for recently developed two-dimensional quantum processors with OBC.
\end{abstract}
\pacs{Valid PACS appear here}
\maketitle

\section{Introduction}

With recent experimental developments in the controllability of noisy intermediate scale quantum (NISQ) computers~\cite{Preskill2018quantumcomputingin,PRXQuantum.2.017001}, more and more attention has been paid to the variational quantum simulation (VQS)~\cite{vqs_a1,vqs1,vqs2,vqs3,vqs4,RevModPhys.94.015004}. One primary task of the VQS is estimating the ground-state energy and preparing the ground state of many-body quantum systems, as target Hamiltonians, via hybrid quantum-classical variational algorithms, which is known as variational quantum eigensolvers (VQEs). A central ingredient of VQEs is the design of parametrized quantum circuits. For instance, a type of VQE, which is based on the hardware efficient ansatz comprised of layers of single-qubit rotation gates and blocks of non-parametrized entangling gates, has been applied to solve the ground-state energy of molecules~\cite{q_chem1,q_chem2,q_chem3,q_chem4,q_chem5} and condensed-matter systems~\cite{q_chem1,BravoPrieto2020scalingof}.

An alternative approach is the VQE inspired by the quantum approximate optimization algorithm (QAOA), which is originally employed to solve combinatorial optimization problems, such as the MaxCut problem~\cite{maxcut_1,maxcut_2,maxcut_3,maxcut_4,maxcut_5,maxcut_a1,2021arXiv211014206B,2022arXiv220203459W}, and the Maximum Independent Set problem~\cite{science_add1}. More recently, numerical works show that the QAOA-inspired VQE can be used to efficiently prepare the ground states of quantum systems~\cite{qaoa_gs1,qaoa_gs2,qaoa_gs3,qaoa_gs4,qaoa_gs5,qaoa_gs7,qaoa_gs6} as well. Experimentally, the ground states of both long-range and short-range Ising models have been generated using the QAOA on trapped-ion quantum simulators~\cite{pnas_1,pnas_2}. Other important applications of the QAOA include the demonstration of quantum advantage~\cite{qaoa_quantum_adv_1,qaoa_quantum_adv_2}, and the generation of the Greenberger-Horne-Zeilinger (GHZ) state~\cite{qaoa_gs1,qaoa_ghz_1}.

Different from the VQE based on the hardware efficient ansatz, the quantum circuit of the QAOA-inspired VQE belongs to a family of problem-inspired ansatz~\cite{RevModPhys.94.015004}. More specifically, it is also called the Hamiltonian variational ansatz~\cite{qaoa_gs2}, because the design of the quantum circuit is closely related to the target Hamiltonian. Moreover, it is worthwhile emphasizing that while there exist barren plateaus on the optimization landscape of both VQEs based on the the hardware efficient ansatz~\cite{barren_plateaus} and the QAOA, the effect of barren plateaus is weaker for the latter~\cite{qaoa_gs2}.



The figure of merit for a VQE boils down to the distance of the VQE's final state from the target state.
Ingredients that affect the performance of the QAOA on the ground-state preparation have been widely explored. It has been shown that essentially, the success of the QAOA depends on the depth of the quantum circuit. Specifically, the ground states of one-dimensional (1D) transverse field Ising model (TFIM) with periodic boundary condition (PBC) and a system size $N$ can be prepared using the QAOA with a perfect fidelity $f=1$, as long as the depth $p$ satisfies $p \ge N/2$~\cite{qaoa_gs1}. Moreover, it has been observed in reference~\cite{qaoa_gs3}, through numerically simulating the ground-state preparation for the non-integrable Ising models, that the perfect fidelity $f=1$ with $p=N/2$ is only attainable for the scenario where the target ground state perturbs near closet free-fermion states.


In this work, we study the performance of QAOA for preparing the ground states of several target Hamiltonians, such as the 1D and two-dimensional (2D) TFIMs with both PBC and open boundary condition (OBC), the Ising model with random interactions and transverse-field strengths, and the 1D Heisenberg model with both PBC and OBC. In contrast to previous works which were mainly restricted to target Hamiltonians with PBC, i.e., translation-invariant systems, our results show that it is harder to perform QAOA without translation-invariant symmetry, for instance due to the OBC or strong randomness. To more efficiently prepare those ground states, we propose a generalized QAOA-inspired VQE assisted by parametrized resource Hamiltonians (PRHs), and numerically demonstrate its advantage through several concrete examples.

In addition to the ground states preparation, we also pay attention to the generation of high-fidelity GHZ states via the QAOA. We find that the PBC is necessary to achieve perfect fidelity of the $N$-qubit GHZ state by using the 1D quantum circuit of the conventional QAOA with a depth $N/2$, and the fidelity significantly drops for the 1D OBC quantum circuit with the same depth. Using the QAOA assisted by the PRH (PRH-QAOA), however, we can also  obtain the perfect fidelity employing the 1D quantum circuit with a depth of $N/2$ even for OBC. Finally, enlightened by the PRH-QAOA, we propose a quantum circuit with a cross-shape geometry beyond 1D, which can generate the $N$-qubit GHZ state with perfect fidelity, requiring a shallower circuit depth than the QAOA using the 1D quantum circuit. We also discuss its experimental realization on Rydberg-dressed atoms.

~\\





\section{Variational quantum eigensolver based on the quantum approximate optimization algorithm} \label{sec2}

\subsection{Quantum circuit and algorithm}

Here, we briefly introduce the quantum circuit and algorithm of the VQE inspired by the conventional QAOA.
The VQE algorithm aims to generate the ground state of a target Hamiltonian $\hat{H}^{(T)}$,
which can be often divided into $M\geq 2$ parts, i.e., $\hat{H}^{(T)}=\sum_{j=1}^{M}\hat{H}^{(T)}_{j}$ with $[\hat{H}^{(T)}_{j},\hat{H}^{(T)}_{j+1}]\neq 0$ and any two terms inside each $\hat{H}^{(T)}_{j}$ commute with each other for all $j$. We define the unitary matrix $U_{j}(x)=\exp(-ix\hat{H}^{(T)}_{j})$, and the quantum circuit of the QAOA-inspired VQE can be realized by repeatedly performing cycles of operations which consist of sequentially switching on and off each term from $\hat{H}^{(T)}_{M}$ to   $\hat{H}^{(T)}_{1}$  as
\begin{eqnarray}
U(\textbf{x}) = \prod_{n=1}^{p} [U_{1}(x_{1}^{(n)})U_{2}(x_{2}^{(n)})\ldots U_{M}(x_{M}^{(n)})]
\label{quantum_circuit}
\end{eqnarray}
with $p$ being the depth of the quantum circuit, and $\textbf{x} = (x_{1}^{(1)},x_{2}^{(1)},\ldots, x_{M}^{(1)},  x_{1}^{(2)}, \ldots,x_{M}^{(p)})$. The parametrized final state of the quantum circuit is $|\psi(\textbf{x})\rangle = U(\textbf{x})|\psi_{0}\rangle$ with $|\psi_{0}\rangle$ being the initial state which can be chosen as the ground state of $\hat{H}^{(T)}_{j}$ ($j\neq M$). Then, the energy $E(\textbf{x}) = \langle\psi(\textbf{x}) |\hat{H}^{(T)}|\psi(\textbf{x})\rangle$ and the fidelity $f(\textbf{x}) = |\langle\psi_{\text{GS}}|\psi(\textbf{x})\rangle|^{2}$, where $|\psi_{\text{GS}}\rangle$ is the ground state of $\hat{H}^{(T)}$, can be directly obtained.

We can generate a final quantum state $|\psi(\textbf{x})\rangle$, which is close to the ground state $|\psi_{\text{GS}}\rangle$ via a hybrid quantum-classical procedure, where the unitary evolution $U(\textbf{x})$ is simulated and the cost function is measured on quantum computers, and the optimization of the parameters $\textbf{x}$ is performed on classical computers. In this work, we adopt the BFGS method, a gradient-based approach, to optimize the cost functions.

It is worthwhile mentioning that except for preparing ground states of target Hamiltonians, QAOA has also been employed to generate multipartite entangled  states, including the GHZ state~\cite{qaoa_gs1,qaoa_ghz_1} and the Dicke state~\cite{Santra2022}. The QAOA may also have the potential to generate other more complex state such as the quantum image~\cite{Zhang2013}, and possibly apply to quantum image detection~\cite{Chetia2021}. In our work, we focus on two basic applications of the QAOA, i.e., the VQE inspired by the QAOA and the generation of high-fidelity GHZ states.



\begin{figure}[]
	\centering
	\includegraphics[width=1\linewidth]{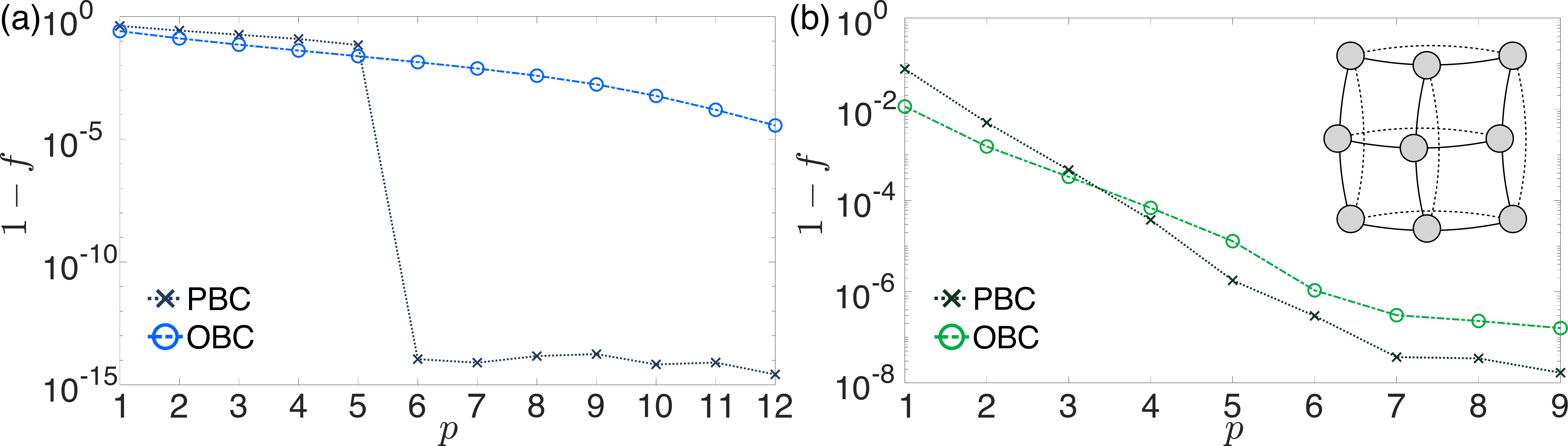}\\
	\caption{(a) The infidelities $1-f$ obtained after optimization as a function of the depth of the VQE $p$ for the 1D ferromagnetic TFIMs with both PBC and OBC. (b) is similar to (a) but for the 2D ferromagnetic TFIMs. The inset of (b) shows a schematic of a regular 2D lattice with $N=9$ spins (represented by circles). For the OBC case, only the interactions represented by solid lines are considered. For the PBC case, both the interactions represented by solid and dashed lines are considered.  }\label{fig1}
\end{figure}



\subsection{Performance of the QAOA-inspired VQE}

We first consider the 1D  ferromagnetic TFIM with $N=12$ as the target Hamiltonian, i.e., $\hat{H}^{(T)} = -\sum_{i=1}^{N}\hat{\sigma}_{i}^{z}\hat{\sigma}_{i+1}^{z} - \lambda \sum_{i=1}^{N}\hat{\sigma}_{i}^{x}$ for PBC ($i+N\equiv i$) and $\hat{H}^{(T)} = -\sum_{i=1}^{N-1}\hat{\sigma}_{i}^{z}\hat{\sigma}_{i+1}^{z} - \lambda \sum_{i=1}^{N}\hat{\sigma}_{i}^{x}$ for OBC. We choose $M=2$. According to equation~(\ref{quantum_circuit}), the  parametrized quantum circuit can be written as
\begin{eqnarray}
U(\textbf{x}) = \prod_{n=1}^{p} [U_{1}(x_{1}^{(n)})U_{2}(x_{2}^{(n)})]
\label{qaoa_ising_conventional}
\end{eqnarray}
with
\begin{eqnarray}
U_{2}(x_{2}^{(n)}) = \exp(i x_{2}^{(n)} \sum_{i=1}^{N(N-1)}\hat{\sigma}_{i}^{z}\hat{\sigma}_{i+1}^{z})
\label{qaoa_ising_conventional_sub1}
\end{eqnarray}
and
\begin{eqnarray}
U_{1}(x_{1}^{(n)}) = \exp(i x_{1}^{(n)} \sum_{i=1}^{N}\hat{\sigma}_{i}^{x}).
\label{qaoa_ising_conventional_sub2}
\end{eqnarray}
The initial state $|\psi_{0}\rangle = \bigotimes_{i=1}^{N} |+\rangle_{i}$  is the ground state of  $\hat{H}^{(T)}_{1} = - \lambda \sum_{i=1}^{N}\hat{\sigma}_{i}^{x}$ ($\lambda>0$), where $|+\rangle_{i}$ is the eigenstate of $\hat{\sigma}_{i}^{x}$ with the eigenvalue $+1$. Here, we focus on the critical point, i.e., $\lambda=1$, separating the ferromagnetic and paramagnetic phase for the TFIM in the thermodynamic limit~\cite{qpt_ising_1d}.


As shown in figure.~\ref{fig1}(a), for the 1D TFIM with PBC, one can achieve a perfect fidelity $f$ ($1-f\simeq 10^{-15}$ to machine precision) with a depth $p\geq N/2$~\cite{qaoa_gs1}. However, for the TFIM with OBC, the perfect fidelity is not achievable with a depth up to $12$, indicating that the performance of the QAOA-inspired VQE is relevant to the boundary condition of the target Hamiltonian.
Next, we study the TFIM on a 2D square lattice with $N=3 \times 3$ sites as the target Hamiltonian, i.e., $\hat{H}^{(T)} = -\sum_{\langle i,j \rangle}\hat{\sigma}_{i}^{z}\hat{\sigma}_{j}^{z} - \lambda \sum_{i=1}^{N}\hat{\sigma}_{i}^{x}$ with $\langle i,j \rangle$ referring to the nearest-neighbor sites $i$ and $j$ on the square lattice [see the inset of figure.~\ref{fig1}(b)]. Similar to the 1D case, we also consider $\lambda=3.05$, as a
critical case near the phase transition point~\cite{qpt_ising_2d,PhysRevE.66.066110}. The results for both OBC and PBC are displayed in figure.~\ref{fig1}(b). It is shown that for both the 1D and 2D models, the infidelities for PBC  are larger than those for OBC when the circuit depth is shallow, whereas it is the opposite for deep circuits.

Moreover, for the 2D TFIM with both PBC and OBC, a perfect fidelity is not achievable. This can be interpreted by the fact that the 2D TFIM can not be mapped to free fermions via the Jordan-Wigner and Bogoloibov transformations~\cite{free_fermion}, and this result generalizes the conjecture of reference~\cite{qaoa_gs3} to 2D systems, suggesting that the perfect fidelity obtained from the QAOA-inspired VQE relies on the target Hamiltonian with a free-fermion-like ground state.



\section{Improvement of the variational quantum eigensolver} \label{sec3}

\subsection{Basic idea}



For the QAOA-inspired VQE, the design of the quantum circuit is closely related to the target Hamiltonian $\hat{H}^{(T)}$. Here, we explore whether the performance of the VQE can be improved if we adopt a resource Hamiltonian $\hat{H}^{(R)}$, which is not identically to the target Hamiltonian. The resource Hamiltonian $\hat{H}^{(R)}$ is then parametrized as $\hat{H}^{(R)} = \hat{H}^{(R)}(\textbf{y})$, where $\textbf{y}=(y_{1},y_{2},\ldots, y_{L})$.
The quantum circuit based on the parametrized resource Hamiltonian (PRH) $\hat{H}^{(R)}$ reads
\begin{eqnarray}
U(\textbf{x},\textbf{y}) = \prod_{n=1}^{p} [U_{1}^{(R)}(x_{1}^{(n)},\textbf{y})\ldots U_{M}^{(R)}(x_{M}^{(n)},\textbf{y})],
\label{quantum_circuit_2}
\end{eqnarray}
where $U_{j}^{(R)}(x,\textbf{y})=\exp[-ix\hat{H}^{(R)}_{j}(\textbf{y})]$, and $\hat{H}^{(R)}(\textbf{y})=\sum_{j=1}^{M}\hat{H}^{(R)}_{j}(\textbf{y})$ with $[\hat{H}^{(R)}_{j}(\textbf{y}),\hat{H}^{(R)}_{j+1}(\textbf{y})]\neq 0$. We note that the parameters $\textbf{y}$ are fixed all the way through the evolution, i.e., $n=1,2,\ldots,p$ in equation~(\ref{quantum_circuit_2}), and so that cannot be absorbed to the parameters $\textbf{x}$.
Next, by optimizing the cost functions, for example the energy and the fidelity, in the parameter space $\textbf{x} \bigcup \textbf{y}$, it is possible to obtain a lower energy or higher fidelity than the conventional QAOA-inspired VQE with $\hat{H}^{(R)}(\textbf{y}) \equiv \hat{H}^{(T)}$. In this paper, the generalized QAOA assisted by the PRH is labeled as PRH-QAOA for short.


Actually, it has been numerically shown that employing the 1D Heisenberg model as the resource Hamiltonian, one can efficiently prepare the ground state of a modified Haldane-Shastry Hamiltonian, providing a revelation as to making the resource Hamiltonian different from the target one in designing QAOA-inspired VQEs~\cite{qaoa_gs2}. We emphasize, however, that in reference~\cite{qaoa_gs2}, the Heisenberg model as the resource Hamiltonian $\hat{H}^{(R)}$ is non-parametrized.
Different from reference~\cite{qaoa_gs2}, here we consider a PRH $\hat{H}^{(R)} = \hat{H}^{(R)}(\textbf{y})$, and will later find out the most appropriate $\hat{H}^{(R)}$ for preparing ground states of other $\hat{H}^{(T)}$.
More technical details of the QAOA-inspired VQE with 1D Heisenberg models being the resource Hamiltonian can be found in Appendix A.

\begin{figure}
	\centering
	\includegraphics[width=1\linewidth]{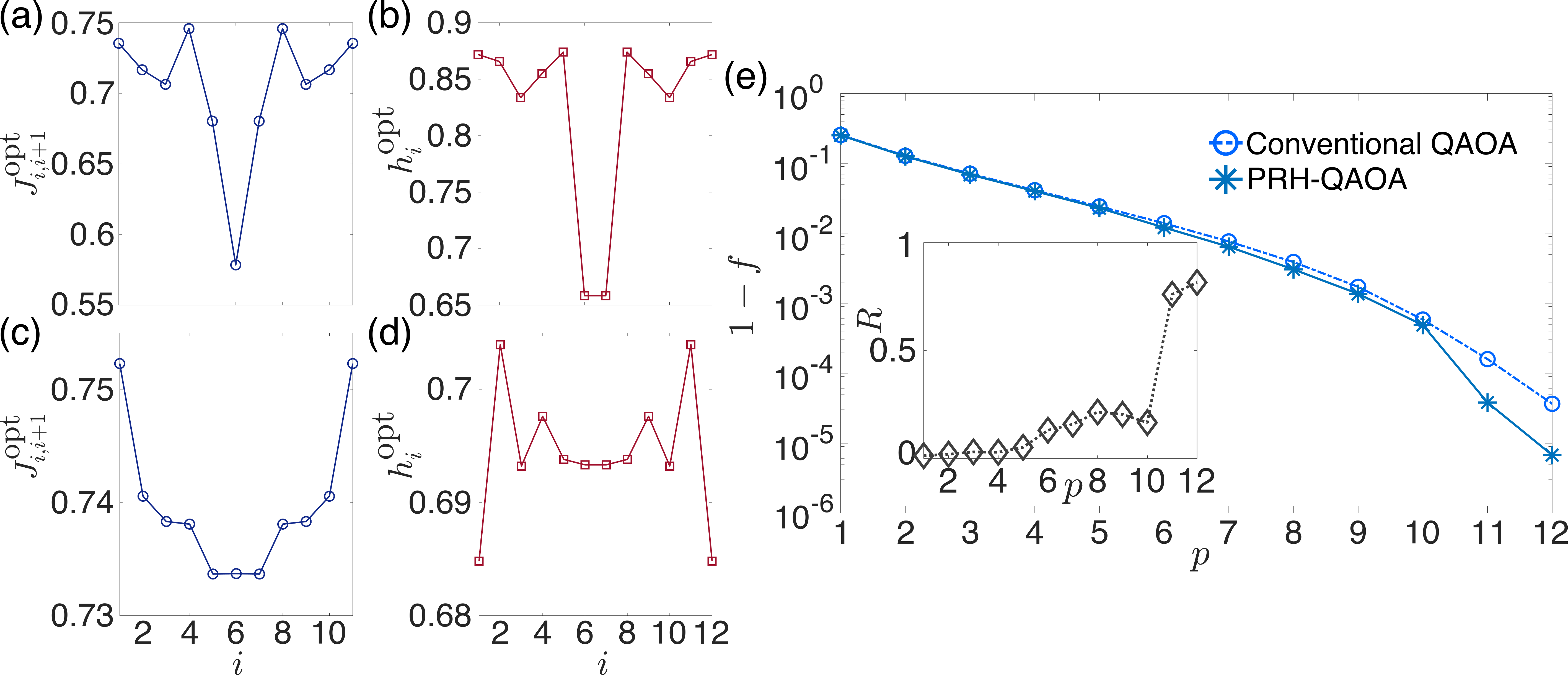}\\
	\caption{(a) The optimized parameter $J^{\text{opt}}_{i,i+1}$ as a function of the spatial index $i$ obtained from the quantum circuit equation~(\ref{quantum_circuit_2}) with a depth $p=6$. (b) The optimized parameter $h^{\text{opt}}_{i}$ as a function of the spatial index $i$ obtained from the quantum circuit equation~(\ref{quantum_circuit_2}) with a depth $p=6$. (c) is similar to (a) but for the quantum circuit with a depth $p=12$. (d) is similar to (b) but for the quantum circuit with a depth $p=12$. (e) The infidelities $1-f$ obtained after optimization as a function of the depth $p$. The inset of (e) shows the ratio $R$ defined by equation~(\ref{ratio}) as a function of the depth $p$. }\label{fig2}
\end{figure}


\subsection{Application 1: one-dimensional transverse-field Ising model with open boundary condition as the target Hamiltonian}


In figure~\ref{fig1}(a), we have seen the performance of conventional QAOA on 1D TFIM with OBC is less efficient. Here, we apply the PRH-QAOA to improve the performance of the VQE for the 1D TFIM with OBC. The target Hamiltonian is still $\hat{H}^{(T)} = -\sum_{i=1}^{N-1}\hat{\sigma}_{i}^{z}\hat{\sigma}_{i+1}^{z} - \sum_{i=1}^{N}\hat{\sigma}_{i}^{x}$ with $N=12$, and we adopt a more generic Ising model with OBC as the resource Hamiltonian, which is divided into $\hat{H}^{(R)}(\textbf{y}) = \hat{H}_{1}^{(R)}(\textbf{h}) + \hat{H}_{2}^{(R)}(\textbf{J})$ with
\begin{eqnarray}
\hat{H}_{1}^{(R)}(\textbf{h}) = -\sum_{i=1}^{N}h_{i} \hat{\sigma}_{i}^{x},
\label{source_1d_ising_1}
\end{eqnarray}
and
\begin{eqnarray}
\hat{H}_{2}^{(R)}(\textbf{J}) = -\sum_{i=1}^{N-1}J_{i,i+1} \hat{\sigma}_{i}^{z}\hat{\sigma}_{i+1}^{z}.
\label{source_1d_ising_2}
\end{eqnarray}
That is we generalize the $\hat{H}^{(R)}$ to allow for parametrized site-dependent interactions and fields therein. It is obvious that the PRH covers the target Hamiltonian, being the $\hat{H}^{(R)}$ in the conventional QAOA, as a subset with $J_{i,i+1}=J$ and $h_i=h, ~\forall j$.
With the above resource Hamiltonian, the parameter space of the quantum circuit~(\ref{quantum_circuit_2}) is $\textbf{x}\bigcup\textbf{y}$ with $\textbf{x}$ being the same as the parameters in equations~(\ref{qaoa_ising_conventional})-(\ref{qaoa_ising_conventional_sub2}), and $\textbf{y} = \textbf{h}\bigcup \textbf{J} = (h_{1},h_{2},\ldots,h_{N})\bigcup (J_{1,2},J_{2,3},\ldots,J_{N-1,N})$.
With $h_{i}>0$ for all $i=1,2,\ldots,N$, the initial state is chosen as $|\psi_{0}\rangle = \bigotimes_{i=1}^{N} |+\rangle_{i}$. Then, we can numerically obtain the final state $|\psi(\textbf{x},\textbf{y})\rangle = U(\textbf{x},\textbf{y})|\psi_{0}\rangle$, and calculate the corresponding cost functions.

After optimizing the fidelity as the cost function, we demonstrate that for the 1D TFIM with OBC, the system with uniform interaction and magnetic field is not the optimal choice. We present two examples of the optimized $\textbf{h}$ and $\textbf{J}$ in figures~\ref{fig2}(a)-(d). It is noted that there exists an inversion symmetry for the optimized parameters, i.e., $J^{\text{opt}}_{i,i+1}\equiv J^{\text{opt}}_{N-i,N-i+1}$ and $h^{\text{opt}}_{i} \equiv h^{\text{opt}}_{N-i+1}$.

In figure~\ref{fig2}(e), we plot the infidelities $1-f$ after optimization for the VQE based on the quantum circuit equation~(\ref{quantum_circuit}) and equation~(\ref{quantum_circuit_2}), known as the conventional QAOA and the PRH-QAOA, respectively.
Moreover, to quantify the improvement of the PRH-QAOA-inspired VQE, we define a ratio
\begin{eqnarray}
R = \frac{(1-f^{C}) - (1-f^{PRH})}{1-f^{C}} = \frac{f^{PRH}-f^{C}}{1-f^{C}}
\label{ratio}
\end{eqnarray}
with $f^{PRH}$ and $f^{C}$ being the fidelity after optimization of the VQE based on the PRH-QAOA and conventional QAOA, respectively, and
plot the value of $R$ as a function of the depth of quantum circuit $p$ in the inset of figure~\ref{fig2}(e). In comparison with the VQE based on the conventional QAOA, one can see the PRH-QAOA-inspired VQE is more efficient for preparing the ground state of the 1D TFIM with OBC.


\subsection{Application 2: two-dimensional transverse-field Ising model as the target Hamiltonian}

\begin{figure}
	\centering
	\includegraphics[width=1\linewidth]{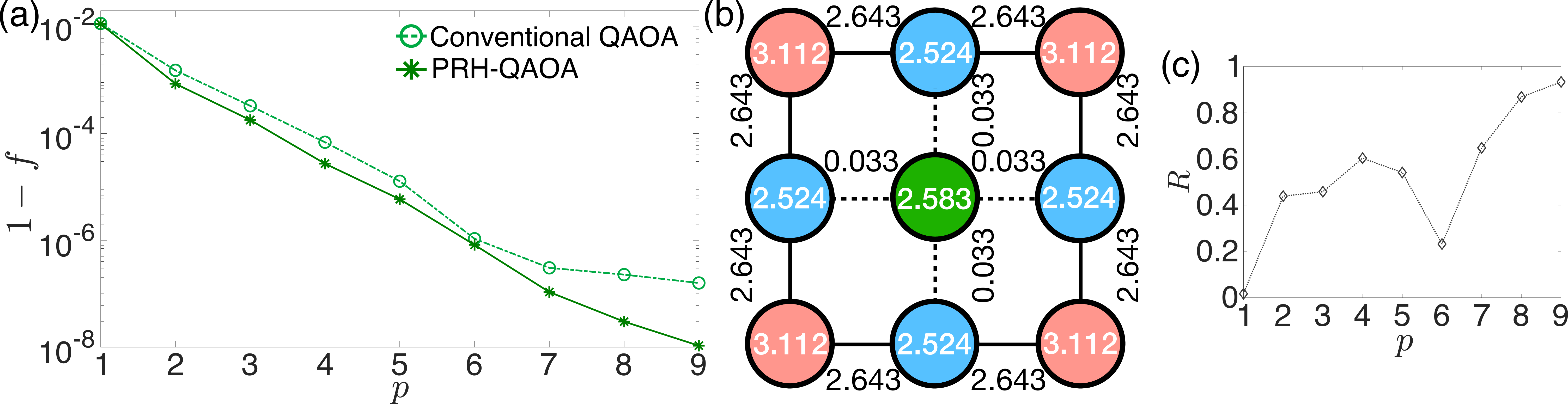}\\
	\caption{(a) The infidelities $1-f$ obtained after optimization as a function of the depth $p$ for the 2D ferromagnetic TFIM. (b) The values of optimized parameters $J^{\text{opt}}_{i,j}$ (near the solid and dashed lines) and $h^{\text{opt}}_{i}$ (inside the circles) with the depth $p=5$. (c) The ratio $R$ defined by equation~(\ref{ratio}) as a function of the depth $p$.  }\label{fig3}
\end{figure}



We now employ the PRH-QAOA-inspired VQE to generate the ground states of 2D ferromagnetic TFIM with $N=9$ and $\lambda=3.05$. Here, we focus on the model with OBC since the open-boundary system is experimentally more relevant for 2D superconducting circuits~\cite{SC_2d_1,SC_2d_2,SC_2d_3} and Rydberg quantum simulators~\cite{rydberg_2d_a1,rydberg_2d_1,rydberg_2d_2,rydberg_2d_3}. We consider the resource Hamiltonian $\hat{H}^{(R)}(\textbf{y}) = \hat{H}_{1}^{(R)}(\textbf{h}) + \hat{H}_{2}^{(R)}(\textbf{J})$ with $\hat{H}_{1}^{(R)}(\textbf{h})$ being the same as equation~(\ref{source_1d_ising_1}) and
\begin{eqnarray}
\hat{H}_{2}^{(R)}(\textbf{J}) = -\sum_{\langle i,j\rangle}J_{i,j} \hat{\sigma}_{i}^{z}\hat{\sigma}_{j}^{z},
\label{source_2d_ising_2}
\end{eqnarray}
which describes ferromagnetic Ising interactions on a 2D square lattice.

As shown in figures~\ref{fig3}(a) and (c), a lower infidelity $1-f$ can be obtained using the VQE based on the PRH-QAOA than the $1-f$ obtained from the conventional QAOA. We present an example of the optimized $\textbf{y}$ for the resource Hamiltonian in figure~\ref{fig3}(b). It is revealed that the optimized $\textbf{y}$ preserves both the inversion symmetry and rotation symmetry.

\subsection{Remarks}


We apply the PRH-QAOA-inspired VQE to the 2D TFIM with PBC and find out that $\hat{H}^{(R)}(\textbf{y}) \equiv \hat{H}^{(T)}$ is the optimal choice, namely generalizing the QAOA to admit site-dependent parameter in the $\hat{H}^{(R)}$ does not help for the target Hamiltonians with PBC.


Additionally, we also consider the Heisenberg model as the target Hamiltonian, and study the performance of both VQE inspired by the conventional QAOA and PRH-QAOA. The results and corresponding discussions are presented in Appendix A. The Heisenberg model is an integrable model exactly solved  using the Bethe ansatz~\cite{qpt_xxz_1d}. However, the Heisenberg model cannot be mapped to free fermions, and thus the perfect fidelity cannot be obtained via the VQE with a $N/2$ depth of quantum circuit.
The dependence of the performance of the conventional QAOA-inspired VQE on the boundary condition is the same as that we have revealed in the TFIMs.
For the Heisenberg model with OBC being the target Hamiltonian, the PRH-QAOA-inspired VQE can also obtain a higher fidelity of the ground state than the conventional QAOA.


Moreover, in Appendix B, we present the results of the energy as the cost function, the 2D TFIM with other values of $\lambda$, as well as the 2D TFIM with triangular geometry,
as the target Hamiltonians.
The additional data displayed in Appendix A and B demonstrate that the improvement of the VQE based on the PRH-QAOA is potential to be generalized to other cases.



\begin{figure}
	\centering
	\includegraphics[width=1\linewidth]{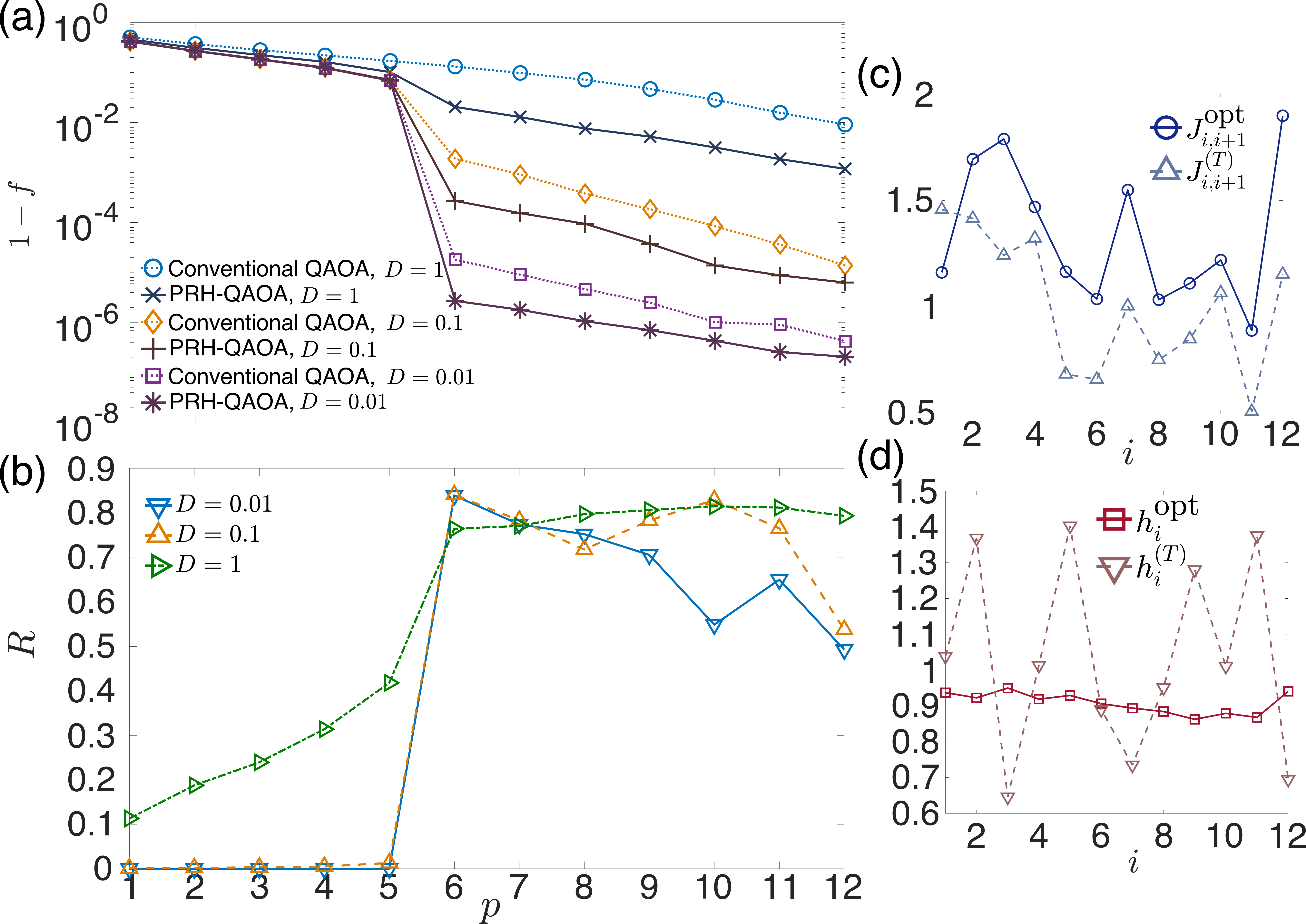}\\
	\caption{(a) The infidelities $1-f$ obtained after optimization as a function of the depth of the VQE $p$ for the randomized Ising models (\ref{rand_ising_tar}). (b) The ratio $R$ defined in equation~(\ref{ratio}) as a function of the depth $p$. (c) An example of the optimized parameter $J_{i,i+1}^{\text{opt}}$ and the corresponding parameter $J_{i,i+1}^{(T)}$ in the target Hamiltonian (\ref{rand_ising_tar}), obtained from the PRH-QAOA-inspired VQE with a depth $p=6$, as a function of the spatial index $i$. (d) is similar to (c) but for the the optimized parameter $h_{i}^{\text{opt}}$ and the corresponding parameter $h_{i}^{(T)}$ in equation~(\ref{rand_ising_tar}). }\label{fig4}
\end{figure}

\section{Generating ground states of Ising models with random interactions and transverse-field strengthes}

After revealing the influence of boundary condition on the performance of QAOA, both for the conventional and the PRH-assisted scenarios, we now consider another source that can break the translational invariance, i.e., the randomness. We focus on the TFIM with PBC as the target Hamiltonian
\begin{eqnarray}
\hat{H}^{(T)} = -\sum_{i=1}^{N} J_{i,i+1}^{(T)} \hat{\sigma}_{i}^{z}\hat{\sigma}_{i+1}^{z} - \sum_{i=1}^{N}h_{i}^{(T)}\hat{\sigma}_{i}^{x},
\label{rand_ising_tar}
\end{eqnarray}
where the Ising interactions $J_{i,i+1}^{(T)}$ and the transverse-field strengthes $h_{i}^{(T)}$ are drawn from a uniform distribution $[(2-D)/2, (2+D)/2]$ with $D$ being the disorder strength. If $D\rightarrow 0$, the Hamiltonian (\ref{rand_ising_tar}) becomes translation-invariant. With the increase of $D$, one can explore how the breakdown of the translational invariance influences the performance of the VQE.

Here, for the conventional QAOA-inspired VQE, the quantum circuit can be written as equation~(\ref{qaoa_ising_conventional}) with $U_{2}(x_{2}^{(n)}) = \exp(ix_{2}^{(n)}\sum_{i=1}^{N}J_{i,i+1}^{(T)} \hat{\sigma}_{i}^{z}\hat{\sigma}_{i+1}^{z})$ and $U_{1}(x_{1}^{(n)}) =\exp(i x_{1}^{(n)}\sum_{i=1}^{N} h_{i}^{(T)}\hat{\sigma}_{i}^{x})$. For the PRH-QAOA-inspired VQE, the quantum circuit can be described by equation~(\ref{quantum_circuit_2}) with the resource Hamiltonian $\hat{H}^{(R)} = \hat{H}_{1}^{(R)}(\textbf{h}) + \hat{H}_{2}^{(R)}(\textbf{J})$, where $\hat{H}_{1}^{(R)}(\textbf{h})$ is the same as equation~(\ref{source_1d_ising_1}), and $\hat{H}_{2}^{(R)}(\textbf{J}) = -\sum_{i=1}^{N} J_{i,i+1} \hat{\sigma}_{i}^{z}\hat{\sigma}_{i+1}^{z}$ with $\textbf{J} = (J_{1,2},J_{2,3},\ldots, J_{12,1})$ for $N=12$. We consider three disorder strengthes $D=1$, $0.1$, and $0.01$.

The results obtained from the VQE based on the conventional QAOA and PRH-QAOA are plotted in figure~\ref{fig4}(a), where the infidelities are obtained by averaging $10$ samples of $J_{i,i+1}^{(T)}$ and $h_{i}^{(T)}$ for each $D$. It is shown that the performance of the VQEs is always better for smaller $D$ regardless of the conventional QAOA or PRH-QAOA, and for each given $D$, the VQE based on PRH-QAOA always outperforms the conventional one. Moreover, as shown in figure~\ref{fig4}(b), when the depth $p\geq 6$, the improvement becomes significant. In figures~\ref{fig4}(c) and (d), we also present an example of optimized parameters in the resource Hamiltonian $\hat{H}^{(R)}$.

\section{Generating the Greenberg-Horne-Zeilinger state via the improved quantum approximate optimization algorithm}


The GHZ states, as a type of multipartite entangled states, are central for numerous quantum-based technologies, ranging from quantum teleportation~\cite{ghz_app1},  quantum metrology~\cite{GHZ_QMetrology} to quantum error-correcting codes~\cite{ghz_app2}. Experimental generation of high-fidelity GHZ states has been achieved on various quantum platforms, including superconducting circuits~\cite{ghz_exp1,ghz_exp4,ghz_exp5,PhysRevA.101.032343}, optomechanical-like quantum systems~\cite{optomech}, quantum optical platforms~\cite{quan_optic_1,quan_optic_2}, trapped ions~\cite{ghz_exp3,PRXQuantum.2.020343} and Rydberg atoms~\cite{ghz_exp2,ghz_exp_rydberg_a}.

In neutral-atom quantum processors, GHZ states with fidelity $f> 0.5$ have been prepared via the optimal control of the adiabatic dynamics~\cite{ghz_exp2,Cui_2017}. In quantum optical platforms, using the Hong-Ou-Mandel interference, multi-photon GHZ states can be generated~\cite{Yao2012}. Digital-gate-model quantum circuits with CNOT gates designed for GHZ states can also be implemented on Rydberg atoms~\cite{ghz_exp_rydberg_a} and superconducting circuits~\cite{Neeley2010}. Nowadays, how to generate high-fidelity and large-scale GHZ states remains a central topic in quantum physics, and
new schemes of preparing multi-qubit GHZ states are timely in need.



\subsection{One-dimensional quantum circuit}

\begin{figure}
	\centering
	\includegraphics[width=1\linewidth]{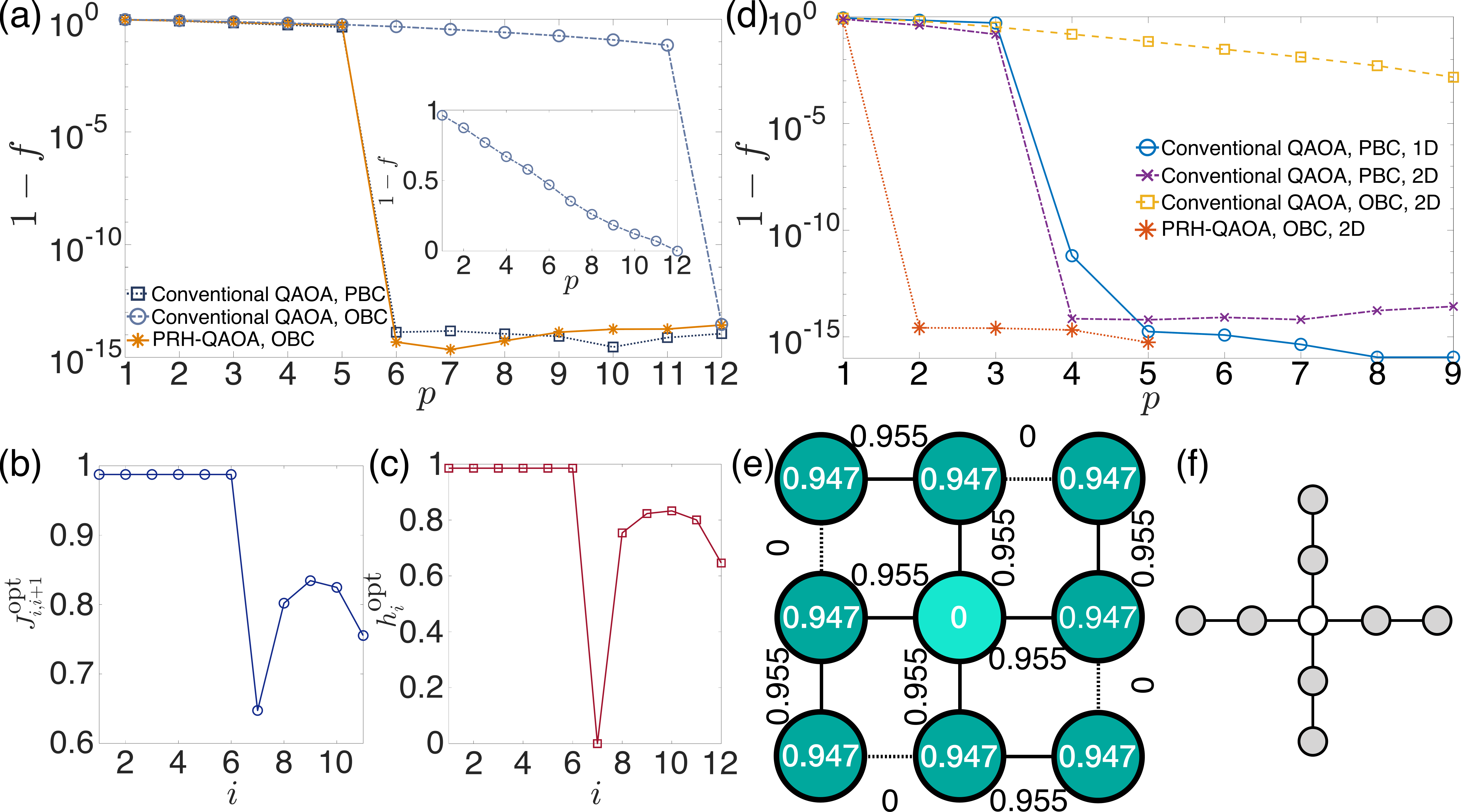}\\
	\caption{(a) The infidelities $1-f$ with $f$ being the GHZ state fidelity, obtained after optimization, as a function of the depth $p$. The resource Hamiltonian is chosen as the 1D ferromagnetic TFIM with OBC or PBC. The inset of (a) plots the $1-f$ as a function of the depth $p$ in a linear y-axis. (b) The optimized parameter $J^{\text{opt}}_{i,i+1}$ as a function of the spatial index $i$ obtained from the PRH-QAOA with a depth $p=6$. (c) is similar to (b) but for the optimized parameter $h^{\text{opt}}_{i}$. (d) The infidelities $1-f$ with $f$ being the GHZ state fidelity, obtained after optimization, as a function of the depth $p$. (e) The values of optimized parameters $J^{\text{opt}}_{i,j}$ (near the solid and dashed lines) and $h^{\text{opt}}_{i}$ (inside the circles) with the depth $p=2$. (f) The geometry of the Ising interactions (presented as the solid lines) of the resource Hamiltonian with $N=9$. The circles in (f) denote the location of qubits. The hollow circles in (c) denote the central qubit. }\label{fig5}
\end{figure}


The GHZ state defined as
\begin{eqnarray}
|\text{GHZ}\rangle \equiv \frac{\bigotimes_{i=1}^{N}|\uparrow\rangle_{i} + \bigotimes_{i=1}^{N}|\downarrow\rangle_{i}}{\sqrt{2}},
\label{ghz_state}
\end{eqnarray}
with $|\uparrow(\downarrow)\rangle_{i}$ being the eigenstate of $\hat{\sigma}_{i}^{z}$ with the eigenvalue $+1(-1)$, is a ground state of the ferromagnetic Ising Hamiltonian $\hat{H}^{(T)} = -\sum_{\langle i,j\rangle} \hat{\sigma}_{i}^{z}\hat{\sigma}_{j}^{z}$. We have demonstrated that the QAOA circuit equations~(\ref{qaoa_ising_conventional}-\ref{qaoa_ising_conventional_sub2}) can generate the ground state of transverse field Ising models. Here, we study the performance of same QAOA circuit $U(\textbf{x}) = \prod_{n=1}^{p} [U_{1}(x_{1}^{(n)})U_{2}(x_{2}^{(n)})]$ with $U_{1}(x_{1}^{(n)}) = \exp(i x_{1}^{(n)} \sum_{i=1}^{N}\hat{\sigma}_{i}^{x})$ and $U_{2}(x_{2}^{(n)}) = \exp(i x_{2}^{(n)} \sum_{i=1}^{N(N-1)}\hat{\sigma}_{i}^{z}\hat{\sigma}_{i+1}^{z})$ for preparing the GHZ state (\ref{ghz_state}), with the cost function
\begin{eqnarray}
f \equiv |\langle \text{GHZ}|\psi(\textbf{x})\rangle|^{2}
\label{f_ghz}
\end{eqnarray}
being the fidelity of the GHZ state.

It has been numerically shown that a $N$-qubit GHZ state with a perfect fidelity $f\simeq 1$ can be generated via the QAOA with the PBC and a depth $p=N/2$~\cite{qaoa_gs1} [also see the results in figure~\ref{fig5}(a)]. Moreover, we show that a perfect fidelity cannot be achieved when employing the OBC in equation~(\ref{qaoa_ising_conventional}). As displayed in figure~\ref{fig5}(a), with the OBC, a 12-qubit GHZ state with a perfect fidelity can be generated if the depth of the QAOA is $p=12$, much deeper than the required depth $p=N/2=6$ when the PBC is adopted.



Next, in order to generate a GHZ state with a perfect fidelity using a shallower depth quantum circuit with OBC, we consider the PRH-QAOA with the circuit
$U_{2}(x_{2}^{(n)}, \mathbf{J}) = \exp(ix_{2}^{(n)}\sum_{i=1}^{N}J_{i,i+1} \hat{\sigma}_{i}^{z}\hat{\sigma}_{i+1}^{z})$, $U_{1}(x_{1}^{(n)},\mathbf{h}) =\exp(i x_{1}^{(n)}\sum_{i=1}^{N} h_{i}\hat{\sigma}_{i}^{x})$, and $U(\textbf{x},\mathbf{J},\mathbf{h}) = \prod_{n=1}^{p} [U_{1}(x_{1}^{(n)},\mathbf{h})U_{2}(x_{2}^{(n)},\mathbf{J})]$, which is the same as the equation~(\ref{quantum_circuit_2}) and the resource Hamiltonian equations~(\ref{source_1d_ising_1}) and (\ref{source_1d_ising_2}). In Appendix C, we display a quantum circuit representation of the aforementioned PRH-QAOA protocol for generating GHZ states. The results of infidelity $1-f$ is plotted in figure~\ref{fig5}(a), and the optimized parameters in the resource Hamiltonian are displayed in figures~\ref{fig5}(b) and (c). It is seen that a 12-qubit GHZ state with $f\simeq 1$ can be generated via the PRH-QAOA with minimal depth $p=6$.


\subsection{Quantum circuit beyond one-dimensional geometry}


\begin{figure}
	\centering
	\includegraphics[width=1\linewidth]{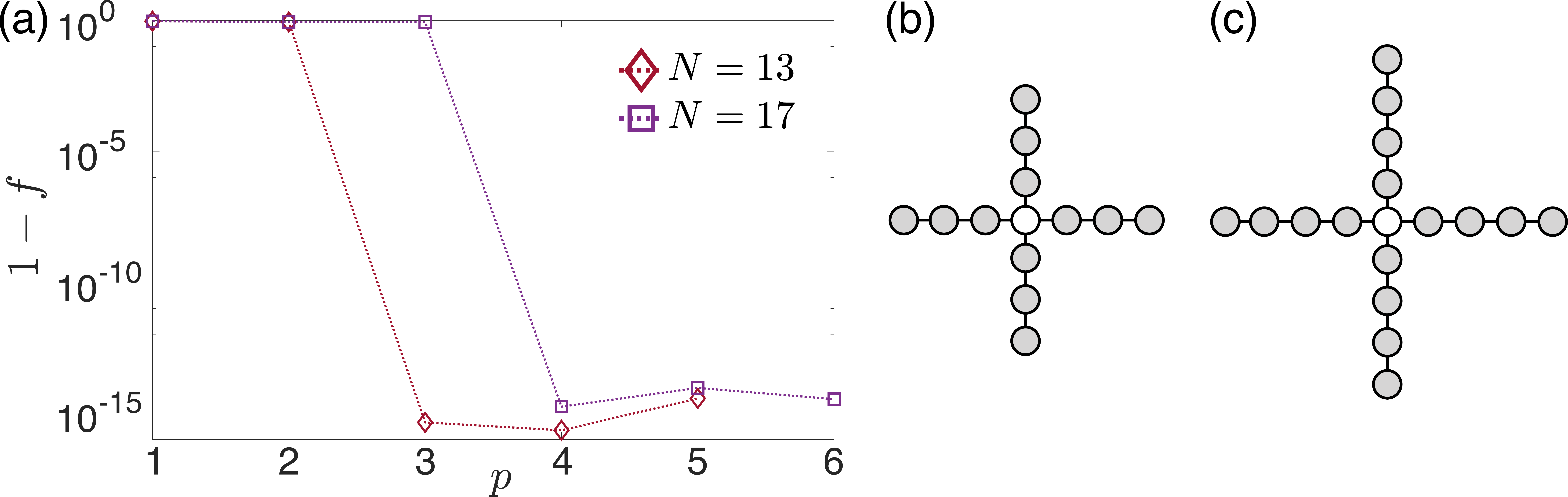}\\
	\caption{(a) The infidelities $1-f$ with $f$ being the GHZ state fidelity, obtained after optimization, as a function of the depth $p$, with the system size $N=13$ and $N=17$. (b) The geometry of the Ising interactions (presented as the solid lines) of the resource Hamiltonian with $N=13$. (c) is similar to (b) but for the system size $N=17$.  }\label{fig7}
\end{figure}


We also study the performance of QAOA for generating GHZ states using a 2D quantum circuit with a size $N=3\times3$, i.e., the TFIM on 2D square lattice as the resource Hamiltonian [see the inset of figure~\ref{fig1}(b)]. We first focus on the conventional QAOA for 2D quantum circuits with both OBC and PBC, in comparison with the results of 1D quantum circuit with PBC. The results of the GHZ state fidelity are shown in figure~\ref{fig5}(d). For a 1D PBC quantum circuit with $N=9$, the fidelity reaches a value $1-f\sim 10^{-11}$ when $p=4$, and a perfect fidelity $1-f\sim 10^{-15}$ can be achieved when $p\geq 5$. For a 2D PBC quantum circuit, a perfect fidelity is attainable when $p\geq 4$. However, with a depth $p\geq 5$, for the 2D OBC quantum circuit, the obtained GHZ state fidelities are much smaller than those for the PBC quantum circuit.

Here, we consider the PRH-QAOA with a 2D OBC quantum circuit. The resource Hamiltonian is the summation of equation~(\ref{source_1d_ising_1}) and equation~(\ref{source_2d_ising_2}). As shown in figure~\ref{fig5}(d), a 9-qubit GHZ state with perfect fidelity can be generated by the algorithm with a quite shallow depth $p=2$ of the quantum circuit. The optimized parameters are presented in figure~\ref{fig5}(e). It is revealed that there are four ferromagnetic Ising interactions whose strength is 0, marked by the dashed lines in figure~\ref{fig5}(e), in the resource Hamiltonian equation~(\ref{source_2d_ising_2}). Moreover, for the resource Hamiltonian equation~(\ref{source_1d_ising_1}), the strength of transverse field of the central qubit is 0.

The resource Hamiltonian with the parameters obtained after optimization leads to a Ising model with a cross-shape geometry of the interactions, as plotted in figure~\ref{fig5}(f). In Appendix C, we also display a quantum circuit representation of the PRH-QAOA protocol with cross-shape geometry for generating GHZ states. The size of the quantum circuit with the cross-shape geometry is scalable, i.e., $N = 5 + 4m$ $(m=0,1,\ldots)$ [see figure~\ref{fig7}(b) and (c)]. Furthermore, the transverse-field strength of the central spin of the resource Hamiltonian is fixed as 0. As shown in figure~\ref{fig7}(a), 13-qubit and 17-qubit GHZ state with perfect fidelity can be generated by the algorithm with the 3-depth and 4-depth quantum circuit, respectively, indicating that employing the Ising model with the special geometry and parameters as the resource Hamiltonian, the QAOA quantum circuit with a depth $m$ can generate $(5 + 4m)$-qubit GHZ state with perfect fidelity.

We note that the shallow depth of the quantum circuit with the ability of generating perfect-fidelity GHZ states benefits from the cross-shape geometry of the resource Ising Hamiltonian plotted in figures~\ref{fig5}(f),~\ref{fig7}(b), and~\ref{fig7}(c). In Appendix D, it is seen that using the square-lattice Ising model with $N=16$, both the conventional QAOA and PRH-QAOA can not generate the perfect-fidelity GHZ state with a 4-depth or shallower depth quantum circuit.

Moreover, we emphasize that all the aforementioned QAOA circuits preparing GHZ states have the global $\mathbb{Z}_{2}$ symmetry given by the parity operator $\hat{P}=\prod_{i=1}^{N}\hat{\sigma}_{i}^{x}$. With the initial state $|\psi_{0}\rangle = \otimes_{i=1}^{N}|+\rangle_{i}$ satisfying $\langle\psi_{0}|\hat{P}|\psi_{0}\rangle=+1$, the symmetry ensures that the final states of the QAOA circuits are in the same symmetry sector $\langle \hat{P}\rangle=+1$. Consequently, the generated ground state of ferromagnetic Ising Hamiltonian $\hat{H}^{(T)}=-\sum_{\langle i,j\rangle}\hat{\sigma}_{i}^{z}\hat{\sigma}_{j}^{z}$ is the GHZ state (\ref{ghz_state}) belonging to the sector $\langle \hat{P}\rangle=+1$, instead of those ground states  out of the sector, such as $\otimes_{i=1}^{N}|\uparrow\rangle_{i}$.




\subsection{Rydberg dressing scheme of realizing the improved QAOA for generating the Greenberg-Horne-Zeilinger state}

\begin{figure*}
	\centering
	\includegraphics[width=1\linewidth]{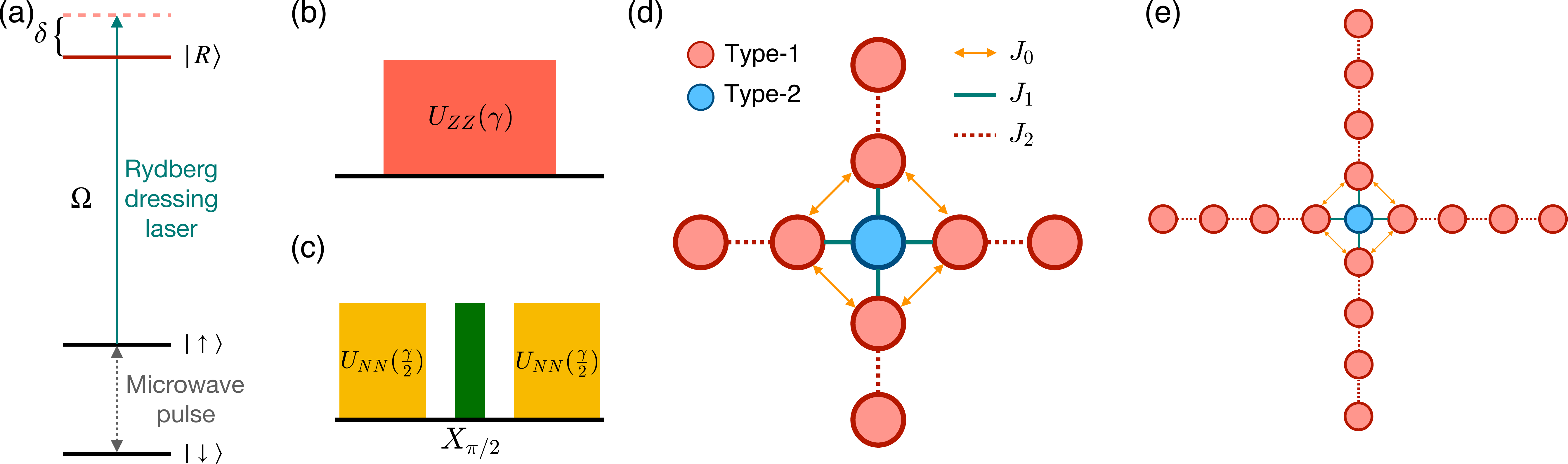}\\
	\caption{(a) Schematic representation of the relevant energy level for a single Rydberg-dressed atom. (b) Schematic representation of Ising dynamics. (c) Schematic representation of the spin echo for realizing Ising dynamics using the unitary dynamics of $\hat{H}_{D}$, i.e., $U_{NN}$, and the single-qubit operation $X_{\pi/2} = \exp[-i(\pi/2)\sum_{i=1}^{N}\hat{\sigma}_{i}^{x}]$. (d) An array of mixed-species atoms beyond 1D geometry enlightened by figures~\ref{fig5}(e) and (f) with $N=9$. Three typical interactions $J_{0}$, $J_{1}$, and $J_{2}$ are relevant. (e) An array of mixed-species atoms with $N=17$.  }\label{fig8}
\end{figure*}


Here, we discuss the experimental realization of generating the GHZ state via the PRH-QAOA with the cross-shape geometry in figures~\ref{fig5}(f),~\ref{fig7}(b) and (c), in Rydberg-dressed neutral atoms~\cite{dressing_1,dressing_2,dressing_3}, accompanied with recent developed mixed-species atom arrays~\cite{mix_atom1,mix_atom2}.

We first consider the generation of the 9-qubit GHZ state.
The realization of this protocol requires an Ising interaction $-\sum_{\langle i,j \rangle} J_{i,j} \hat{\sigma}^{z}_{i}\hat{\sigma}^{z}_{j} $
with $\langle i,j \rangle$ meaning the nearest-neighbor sites $i$ and $j$ on the lattice in figure~\ref{fig5}(f).
For Rydberg-dressed quantum processors, qubits are encoded in two hyperfine ground states $|\downarrow\rangle$ and $|\uparrow\rangle$, with one of the ground state $|\uparrow\rangle$ coupled to a Rydberg state $|R\rangle$ via a laser field of detuning $\delta$ and Rabi frequency $\Omega$~\cite{dressing_1,dressing_2,dressing_3} [also see figure~\ref{fig8}(a)]. When $\Omega/\delta \ll 1$, the Rydberg-dressed state can induce a pairwise energy shift dependent on the distance between two atoms, which can be described by the Hamiltonian
\begin{eqnarray}
\hat{H}_{D} = \sum_{\langle i,j \rangle} V_{i,j} \hat{n}_{i}\hat{n}_{j},
\label{Dressing_Hzz}
\end{eqnarray}
where $\hat{n}_{i} = |\uparrow\rangle_{i} \langle\uparrow |_{i}$, $\hat{n} = (\hat{\sigma}^{z}+\textbf{1})/2$ with $\textbf{1}$ being the two-dimensional identity matrix, and the strength of Rydberg-dressed interaction reads
\begin{eqnarray}
V_{i,j} = V_{0}\frac{R_{C}^{6}}{|\textbf{r}_{i} - \textbf{r}_{j}|^{6} + R_{C}^{6}}
\label{Dressing_int}
\end{eqnarray}
with the positions of atoms $\textbf{r}_{i,j}$, $V_{0} = \hbar\Omega^{4}/8\delta^{3}$, $R_{C}=|C_{6}/2\hbar\delta|^{1/6}$, and $C_{6}$ being the van der Waals coefficient of the Rydberg state. We define the unitary dynamics of $\hat{H}_{D}$ as $U_{NN}(t) = \exp(-it\hat{H}_{D})$. By using a spin echo~\cite{dressing_1,dressing_2} shown in figure~\ref{fig8}(c), one can realize the Ising dynamics [see figure~\ref{fig8}(b)]
\begin{eqnarray}
U_{ZZ}(\gamma) = \exp[-i\gamma \sum_{\langle i,j\rangle} (V_{i,j}/4)\hat{\sigma}^{z}_{i}\hat{\sigma}^{z}_{j}].
\label{Ising_dynamics}
\end{eqnarray}

According to equation~(\ref{Dressing_int}), non-local interactions should be considered. Here, we take into account the additional interactions presented as $J_{0}$ in figure~\ref{fig8}(d). The nearest-neighbor interactions includes $J_{1}$ and $J_{2}$ in figure~\ref{fig8}(d). Both the interactions $J_{1}$ and $J_{2}$ can be adjusted by changing the corresponding intra-atom distance. In addition to the distance, the interaction $J_{1}$ can also be adjusted by choosing different species type-2 [see figure~\ref{fig8}(d)] and different Rydberg states, which essentially results in different coefficient $C_{6}$.
In short, the interactions $J_{0}$, $J_{1}$, and $J_{2}$ can be separately adjusted. Moreover, the Rydberg dressing quantum processor with mixed-species atoms~\cite{mix_atom1,mix_atom2} allows the single-qubit gate performed on the central qubit, i.e., the type-2 atom in figure~\ref{fig8}(d), and a different single-qubit operation on other qubits [the type-1 atom in figure~\ref{fig8}(d)] except for the central one.


Enlightened by the results shown in figure~\ref{fig5}(e), for the QAOA, the resource Hamiltonian is chosen as
\begin{eqnarray}
\hat{H}^{(R)}_{1} = - \sum_{i\in \{\text{type-1}\}} \hat{\sigma}_{i}^{x},
\label{source_rydberg_1}
\end{eqnarray}
and
\begin{eqnarray}
\hat{H}^{(R)}_{2}(J_{0},J_{1},J_{2}) = -\sum_{\langle i,j\rangle} J_{i,j} \hat{\sigma}_{i}^{z}\hat{\sigma}_{j}^{z},
\label{source_rydberg_2}
\end{eqnarray}
with $J_{i,j}\in \{J_{0},J_{1},J_{2}\}$. Based on equation~(\ref{quantum_circuit_2}), the quantum circuit with a depth $p=2$ can be written as
\begin{eqnarray}
U(\textbf{x},\textbf{y}) = U_{1}^{(R)}(x_{1}^{1})U_{2}^{(R)}(x_{2}^{1},\textbf{y}) U_{1}^{(R)}(x_{1}^{2})U_{2}^{(R)}(x_{2}^{2},\textbf{y})
\label{qc_aaa}
\end{eqnarray}
with $\textbf{x} = (x_{1}^{1},x_{2}^{1},x_{1}^{2},x_{2}^{2})$ and $\textbf{y} = (J_{0},J_{1},J_{2})$. The initial state is $|\psi_{0}\rangle = \otimes_{i=1}^{N}|+\rangle_{i}$. The preparation of the initial state $|\psi_{0}\rangle$ and the unitary evolution $U_{1}^{(R)}(x)$ can be experimentally realized by imposing the microwave pulse on the qubits encoded by the hyperfine ground state $|\uparrow\rangle$ and $|\downarrow\rangle$ of neutral atoms~\cite{dressing_1,dressing_2} [also see figure~\ref{fig8}(a)]. The unitary evolution $U_{2}^{(R)}(x,\textbf{y})$ is the same as equation~(\ref{Ising_dynamics}), which can be realized by the dressing pulse and the spin echo [see figures~\ref{fig8}(a)-(c)]. After optimization, we obtain the parameters for generating the GHZ state with perfect fidelity, i.e., $\textbf{y}^{\text{opt}} = (4/3,1,1/3)$ and $\textbf{x}^{\text{opt}} = (\pi/4,3\pi/4,3\pi/4,3\pi/4)$.



Finally, we generalize the results of $N=9$ to larger size $N=13$ and $17$ [see figure.~\ref{fig8}(e) for the geometry of the lattice with $N=17$]. Generating the perfect-fidelity GHZ with $N=13$ and $17$ requires the depth of the quantum circuit $p=3$, and $p=4$, respectively. For $N=13$ and $p=3$, the optimized parameters are $\textbf{y}^{\text{opt}} = (4/3,1,1/3)$ and $\textbf{x}^{\text{opt}} = (x_{1}^{1},x_{2}^{1},x_{3}^{1},x_{1}^{2},x_{2}^{2},x_{3}^{2}) = (\pi/4,\pi/4,3\pi/4,3\pi/4,3\pi/4,3\pi/4)$. Moreover, for $N=17$ and $p=4$, $\textbf{y}^{\text{opt}} = (4/3,1,1/3)$ and $\textbf{x}^{\text{opt}} = (x_{1}^{1},x_{2}^{1},x_{3}^{1},x_{4}^{1},x_{1}^{2},x_{2}^{2},x_{3}^{2},x_{4}^{2} )= (\pi/4,\pi/4,\pi/4,3\pi/4,3\pi/4,3\pi/4,3\pi/4,3\pi/4)$.

\section{Discussion}


We have studied the performance of the QAOA-inspired VQE for target Hamiltonians with and without translation-invariant symmetry. For systems without the symmetry, including the models with OBC or strong randomness, the performance of the VQE is less efficient, in comparison with the case of translation-invariant target Hamiltonians. In general, QAOA steers a controllable quantum system, which we call the resource Hamiltonian in this paper. Previous studies in the literature mainly focused on the QAOA-inspired VQE, restricted to the scenario that the resource Hamiltonian is exactly identical to the target Hamiltonian. We generalize the QAOA to allow the PRH to be possibly different from the target Hamiltonian. If the translation-invariant symmetry in the target Hamiltonian is broken, one can use the PRH-QAOA-inspired VQE to achieve higher fidelity of the target ground state or lower energy corresponding to the target Hamiltonian. Our work provides an essential step toward an application of the PRH-QAOA to preparing ground states of more generic models. An example is the quantum Sherrington-Kirkpatrick (QSK) model with the all-to-all random Ising interactions~\cite{QSK2,QSK1,QSK3,QSK4}. Obviously, the translation-invariant symmetry is absent for the QSK model, and the performance of PRH-QAOA for generating the ground states of the QSK model is worthy of future considerations.  


Except for the QAOA, shortcuts to adiabaticity, being the fast routes to the final results of quantum annealing, is another typical way of obtaining ground states of target Hamiltonians~\cite{STA1,STA2}. However, near the quantum critical point, the adiabatic state preparation is less efficient~\cite{STA3}. An interesting future task is to study whether the PRH-QAOA outperforms the shortcuts to adiabaticity for preparing quantum critical states.

In the NISQ era, quantum computers are not perfectly translation-invariant, due to their OBC~\cite{maxcut_4,SC_2d_1,SC_2d_2,SC_2d_3,rydberg_2d_a1,rydberg_2d_1,rydberg_2d_2,rydberg_2d_3,SC_1d_1,SC_1d_2,SC_1d_3,SC_1d_4}, and randomness induced by experimental imperfections. Consequently, the PRH-QAOA-inspired VQE proposed in our work opens up an avenue for the VQE experimentally generating states closer to the target ground state than using the conventional protocol. Moreover, it is also shown that the perfect fidelity GHZ state can be generated by the PRH-QAOA with shallow-depth quantum circuit. This protocol is experimentally accessible because the affordable circuit depth is limited for lack of efficient error correction technique for NISQ computers. We also briefly discuss the effect of decoherence on the PRH-QAOA generating the GHZ state in Appendix E.
Taking the non-local interactions in Rydberg-dressed quantum simulators~\cite{dressing_2} into consideration, we believe our work provides an alternative way of generating high-fidelity GHZ states in neutral atoms.

Our work points to several further directions: (i) exploring how the translation-invariant symmetry of target Hamiltonians or quantum circuits can influence the performance of VQEs based on other ansatz, such as the hardware efficient ansatzes~\cite{q_chem1,q_chem2,q_chem3,q_chem4,q_chem5,BravoPrieto2020scalingof}, (ii) exploring the performance of QAOA-inspired VQEs for other complex target Hamiltonians, such as trapped-ion systems with long-range interactions satisfying a power-law decay~\cite{vqs1,pnas_1,pnas_2,RevModPhys.93.025001}, and (iii) finding more efficient ansatzes, taking the PRH into consideration, to design quantum circuits of VQEs, in combination with other QAOA-like subsequences~\cite{qaoa_qa1,qaoa_qa2,qaoa_improve1,qaoa_improve2,qaoa_improve3}, for example, the counterdiabatic QAOA~\cite{qaoa_gs6,cd_qaoa_a1,cd_qaoa_a2}.




\begin{acknowledgments}

We acknowledge the discussions with Roeland Wiersema, Kunpeng Wang, and Cheng Sheng. This work was supported by the National Key Research and Development Program of China (Grant No. 2021YFA1402001), the National Science Foundation of China (NSFC) (Grant Nos. T2121001, 11934018, 11904018), Strategic Priority Research Program of Chinese Academy of Sciences (Grant No. XDB28000000), Beijing Natural Science Foundation (Grant No. Z200009), and Scientific Instrument Developing Project of Chinese Academy of Sciences (Grant No. YJKYYQ20200041).




\end{acknowledgments}




\appendix

\section{Results for 1D Heisenberg models}

\begin{figure}
	\centering
	\includegraphics[width=0.9\linewidth]{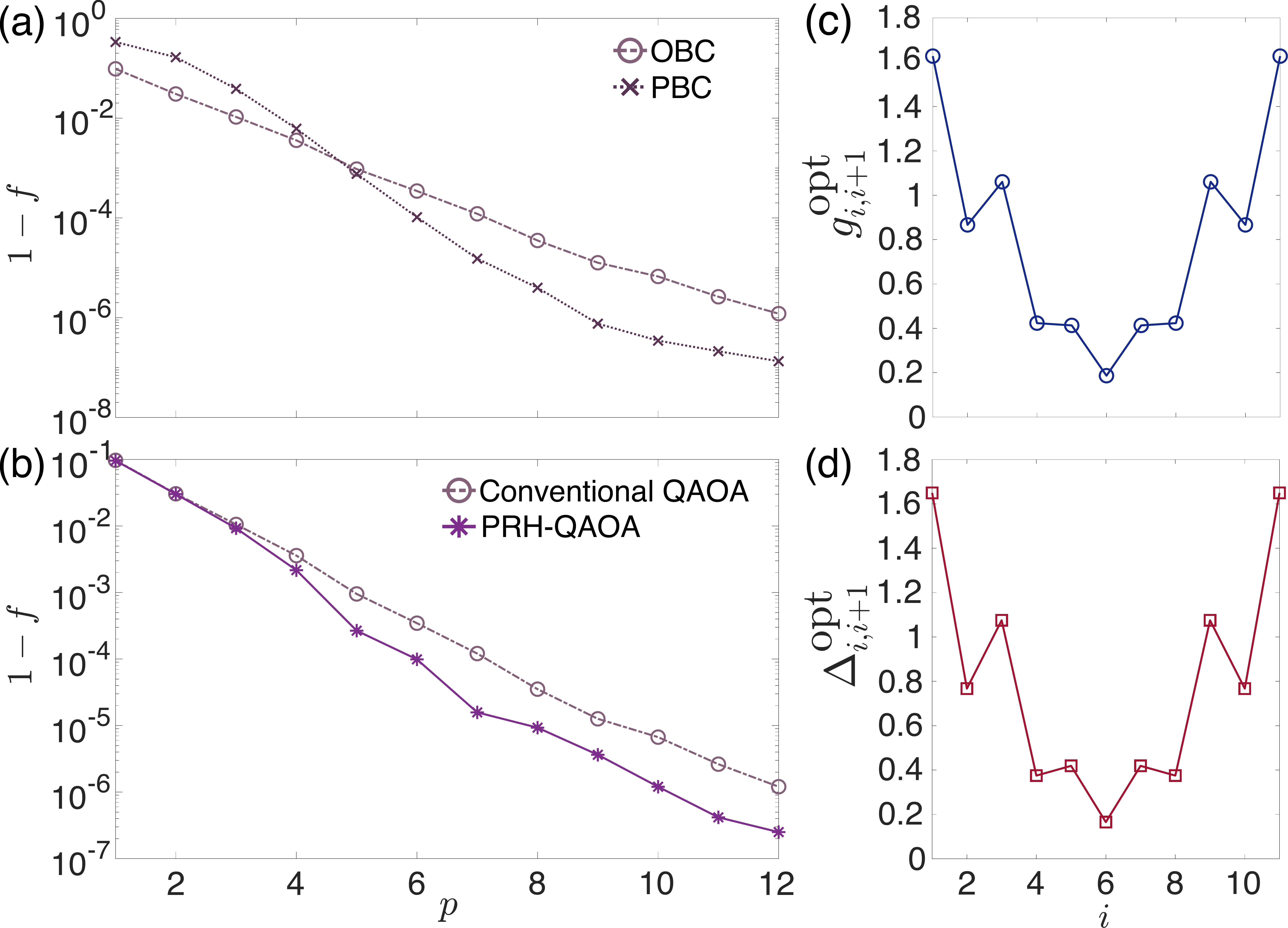}\\
	\caption{(a) The infidelities $1-f$ obtained after optimization as a function of the depth $p$ for the Heisenberg model with both PBC and OBC, using the conventional QAOA-inspired VQE. (b) The infidelities $1-f$ obtained after optimization as a function of the depth $p$ for the Heisenberg model with OBC, using the VQE based on conventional QAOA and PRH-QAOA. (c) The optimized parameter $g^{\text{opt}}_{i,i+1}$ as a function of the spatial index $i$ obtained from the PRH-QAOA with a depth $p=6$. (d) is similar to (c) but for the optimized parameter $\Delta^{\text{opt}}_{i}$.}\label{fig_a1}
\end{figure}

The Hamiltonian of the 1D Heisenberg model with PBC is given by
\begin{eqnarray}
\hat{H} = \sum_{i=1}^{N} [\hat{\sigma}^{x}_{i}\hat{\sigma}^{x}_{i+1} + \hat{\sigma}^{y}_{i}\hat{\sigma}^{y}_{i+1} + \hat{\sigma}^{z}_{i}\hat{\sigma}^{z}_{i+1}].
\label{Heisenberg}
\end{eqnarray}
For the model with OBC, the superscript of the summation in equation~(\ref{Heisenberg}) is replaced by $N-1$. The system size is $N=12$. To prepare the ground states of the 1D Heisenberg model with both OBC and PBC, we design the quantum circuit of the QAOA-inspired VQE by employing equation~(\ref{Heisenberg}) with corresponding boundary conditions being the resource Hamiltonian. According to the VQE based on the conventional QAOA in reference~\cite{qaoa_gs2}, the resource Hamiltonian (\ref{Heisenberg}) is divided into four parts, i.e., $\hat{H}^{\text{even}}_{XY} + \hat{H}^{\text{even}}_{Z} + \hat{H}^{\text{odd}}_{XY} + \hat{H}^{\text{odd}}_{Z}$, where $\hat{H}^{\text{even}}_{XY} = \sum_{i=1}^{N/2}[\hat{\sigma}^{x}_{2i-1}\hat{\sigma}^{x}_{2i} + \hat{\sigma}^{y}_{2i-1}\hat{\sigma}^{y}_{2i}]$, $\hat{H}^{\text{odd}}_{XY} = \sum_{i=1}^{N/2}[\hat{\sigma}^{x}_{2i}\hat{\sigma}^{x}_{2i+1} + \hat{\sigma}^{y}_{2i}\hat{\sigma}^{y}_{2i+1}]$, $\hat{H}^{\text{even}}_{Z} = \sum_{i=1}^{N/2}\hat{\sigma}^{z}_{2i-1}\hat{\sigma}^{z}_{2i}$, and $\hat{H}^{\text{odd}}_{Z} = \sum_{i=1}^{N/2}\hat{\sigma}^{z}_{2i}\hat{\sigma}^{z}_{2i+1}$. According to equation~(\ref{quantum_circuit}), the quantum circuit can be written as
\begin{eqnarray}
U(\textbf{x}) = \prod_{n=1}^{p} [U_{1}(x_{1}^{(n)})U_{2}(x_{2}^{(n)}) U_{3}(x_{3}^{(n)})U_{4}(x_{4}^{(n)})]
\label{qc_Heisenberg}
\end{eqnarray}
with $U_{1}(x_{1}^{(n)}) = \exp(-ix_{1}^{(n)}\hat{H}^{\text{even}}_{Z})$, $U_{2}(x_{2}^{(n)}) = \exp(-ix_{2}^{(n)}\hat{H}^{\text{even}}_{XY})$, $U_{3}(x_{3}^{(n)}) = \exp(-ix_{3}^{(n)}\hat{H}^{\text{odd}}_{Z})$, and $U_{4}(x_{4}^{(n)}) = \exp(-ix_{4}^{(n)}\hat{H}^{\text{odd}}_{XY})$. The initial state of the VQE is chosen as the ground state of both $\hat{H}^{\text{even}}_{XY}$ and $\hat{H}^{\text{even}}_{Z}$, i.e., $|\psi_{0}\rangle = \bigotimes_{i=1}^{N/2} (|0\rangle_{2i-1}|1\rangle_{2i} - |1\rangle_{2i-1}|0\rangle_{2i})/\sqrt{2}$.

In reference~\cite{qaoa_gs2}, the quantum circuit (\ref{qc_Heisenberg}) has been employed to prepare the ground state of the modified Haldane-Shastry Hamiltonian $H^{(T)} = \sum_{n<m}^{N}\frac{1}{d_{nm}^{2}}[\hat{\sigma}^{z}_{n}\hat{\sigma}^{z}_{m} -\hat{\sigma}^{x}_{n}\hat{\sigma}^{x}_{m} - \hat{\sigma}^{y}_{n}\hat{\sigma}^{y}_{m}]$ with a complex long-range interactions $d_{nm} = \frac{N}{\pi}|\sin[\frac{\pi}{N}(n-m)]|$. Here, the target Hamiltonian $H^{(T)}$ is different from the resource Hamiltonian (\ref{Heisenberg}) where the interactions are nearest-neighbor.

As shown in figure~\ref{fig_a1}(a), the boundary condition of the 1D Heisenberg model still influences the performance of the conventional QAOA-inspired VQE, similar to the results in figure~\ref{fig1}. To more efficiently prepare the ground state of the 1D Heisenberg model with OBC, we also consider the quantum circuit of the PRH-QAOA-inspired VQE according to equation~(\ref{quantum_circuit_2}). The PRH is
\begin{eqnarray}
\hat{H}^{(R)}(\textbf{y}) = \sum_{i=1}^{N-1} [g_{i,i+1}(\hat{\sigma}^{x}_{i}\hat{\sigma}^{x}_{i+1} + \hat{\sigma}^{y}_{i}\hat{\sigma}^{y}_{i+1})
+ \Delta_{i,i+1} \hat{\sigma}^{z}_{i}\hat{\sigma}^{z}_{i+1}],
\label{Heisenberg_source}
\end{eqnarray}
which is decomposed into four parts, i.e., $\hat{H}^{(R)}(\textbf{y}) = \hat{H}^{(R)}_{1}(\textbf{y}_{1})+\hat{H}^{(R)}_{2}(\textbf{y}_{2})+\hat{H}^{(R)}_{3}(\textbf{y}_{3})+\hat{H}^{(R)}_{4}(\textbf{y}_{4})$ with $\hat{H}^{(R)}_{1}(\textbf{y}_{1}) = \sum_{i=1}^{N/2}g_{2i-1,2i}[\hat{\sigma}^{x}_{2i-1}\hat{\sigma}^{x}_{2i} + \hat{\sigma}^{y}_{2i-1}\hat{\sigma}^{y}_{2i}]$,
$\hat{H}^{(R)}_{2}(\textbf{y}_{2}) = \sum_{i=1}^{N/2}\Delta_{2i-1,2i}\hat{\sigma}^{z}_{2i-1}\hat{\sigma}^{z}_{2i}$, $\hat{H}^{(R)}_{3}(\textbf{y}_{3}) = \sum_{i=1}^{N/2}g_{2i,2i+1}[\hat{\sigma}^{x}_{2i}\hat{\sigma}^{x}_{2i+1} + \hat{\sigma}^{y}_{2i}\hat{\sigma}^{y}_{2i+1}]$, and $\hat{H}^{(R)}_{4}(\textbf{y}_{4}) = \sum_{i=1}^{N/2}\Delta_{2i,2i+1}\hat{\sigma}^{z}_{2i}\hat{\sigma}^{z}_{2i+1}$. In comparison with the conventional QAOA-inspired VQE, the additional parameters in the resource Hamiltonian are $\textbf{y} = \textbf{y}_{1} \bigcup\textbf{y}_{2}\bigcup \textbf{y}_{3} \bigcup\textbf{y}_{4}$, with $\textbf{y}_{1} = (g_{1,2},g_{3,4},\ldots,g_{11,12})$, $\textbf{y}_{2} = (\Delta_{1,2},\Delta_{3,4},\ldots,\Delta_{11,12})$, $\textbf{y}_{3} = (g_{2,3},g_{4,5},\ldots,g_{10,11})$, and $\textbf{y}_{4} = (\Delta_{2,3},\Delta_{4,5},\ldots,\Delta_{10,11})$. The results of the PRH-QAOA-inspired VQE are plotted in figures~\ref{fig_a1}(b)-(d). Similar to the results in figures~\ref{fig2}(a)-(d), the parameters of the resource Hamiltonian $\textbf{y}$ obtained from optimization still satisfy an inversion symmetry [see figures~\ref{fig_a1}(c) and (d)].

\section{Additional numerics of Ising models}

\begin{figure}
	\centering
	\includegraphics[width=0.6\linewidth]{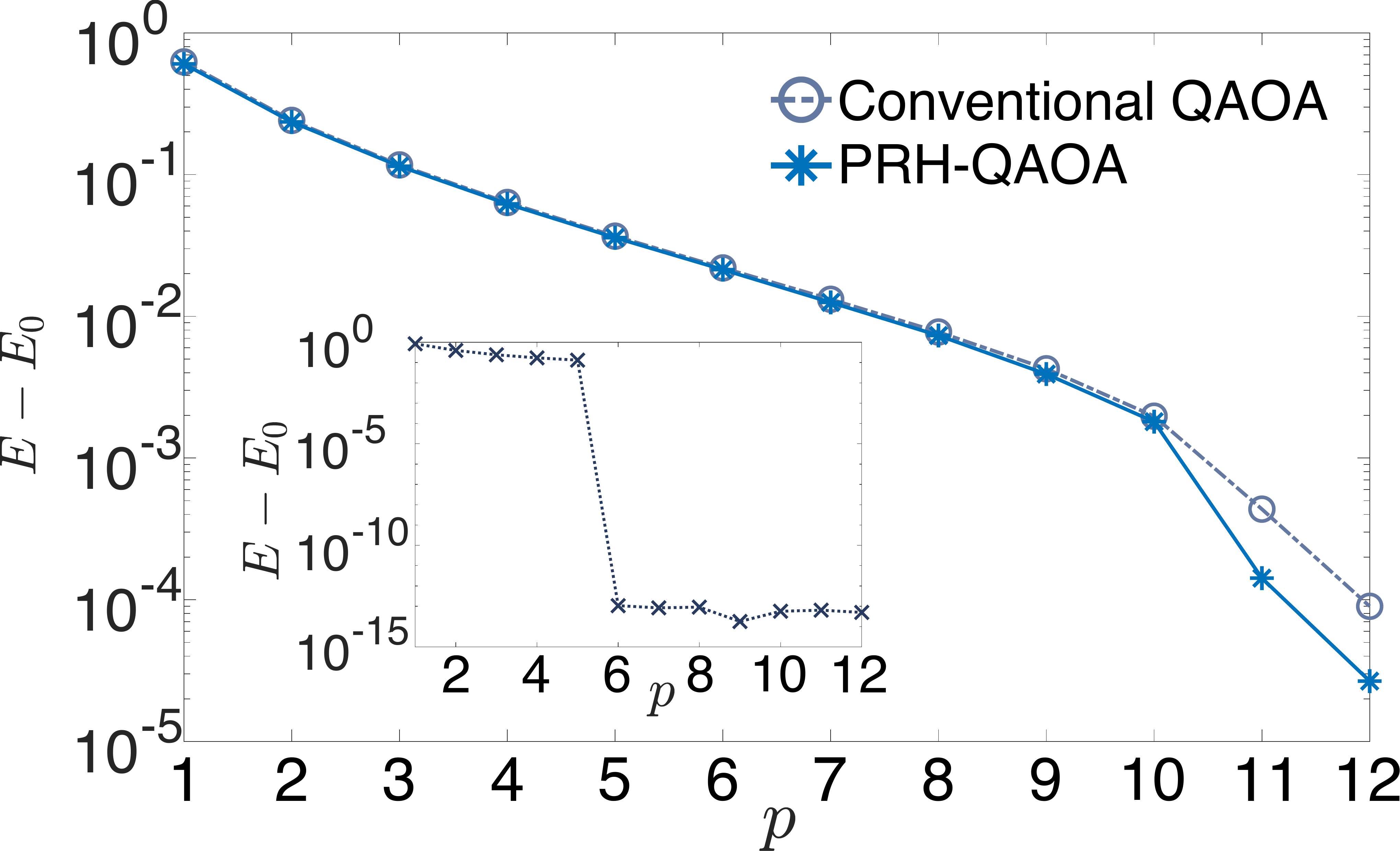}\\
	\caption{The energy difference $E-E_{0}$ obtained after optimization as a function of the depth of the VQE $p$ for the 1D ferromagnetic TFIM with OBC, where $E_{0}$ is the ground state energy of target Hamiltonians. The inset shows the energy difference $E-E_{0}$ as a function of $p$ for the TFIM with PBC.}\label{fig_b1}
\end{figure}

\begin{figure}
	\centering
	\includegraphics[width=1\linewidth]{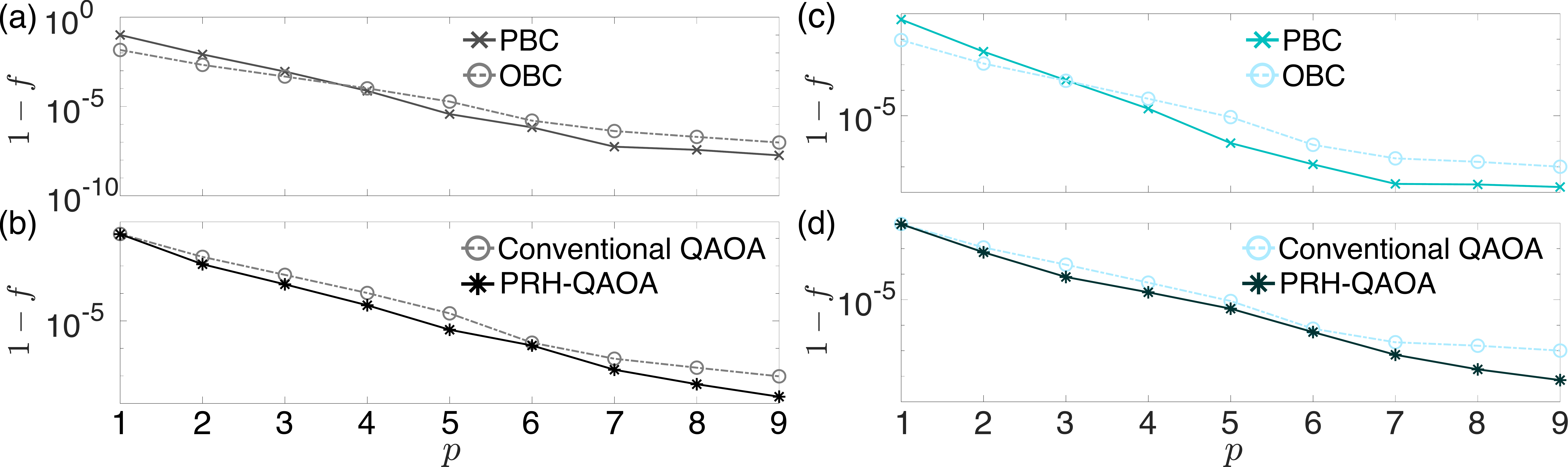}\\
	\caption{(a) The infidelities $1-f$ obtained after optimization using the conventional QAOA-inspired VQE, as a function of the depth $p$ for the 2D ferromagnetic TFIM with both PBC and OBC, and a transverse-field strength $\lambda = 2.9$. (b) The infidelities $1-f$ obtained after optimization using the VQE based on both the conventional QAOA and PRH-QAOA, as a function of the depth $p$ for the 2D ferromagnetic TFIM with OBC. (c) is similar to (a), but for a transverse-field strength $\lambda = 3.2$. (d) is similar to (b), but for a transverse-field strength $\lambda = 3.2$.}\label{fig_b2}
\end{figure}

\begin{figure}
	\centering
	\includegraphics[width=1\linewidth]{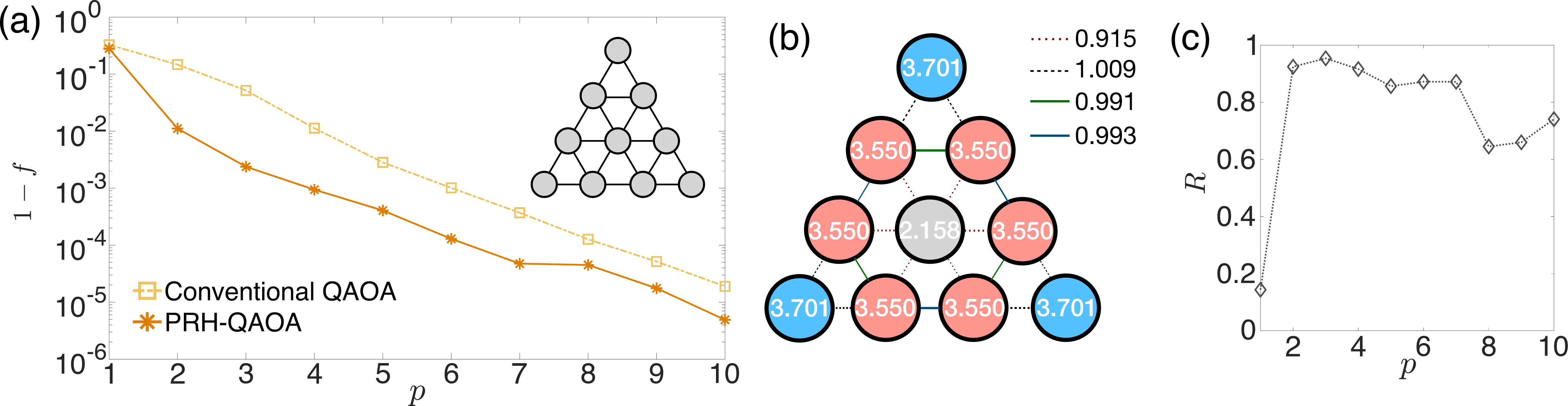}\\
	\caption{(a) The infidelities $1-f$ obtained after optimization as a function of the depth $p$ for the 2D ferromagnetic TFIM with the triangular geometry. (b) The values of optimized parameters $J_{i,j}^{\text{opt}}$ (represented by dashed lines, dotted lines, and solid lines with different colors) and $h_{i}^{\text{opt}}$ (inside the circles) with the depth $p=6$. (c) The ratio $R$ defined by equation~(\ref{ratio}) as a function of $p$. The inset of (a) shows a schematic of a triangular 2D lattice with $N=10$ spins (represented by circles), and the ferromagnetic Ising interactions between spins (represented by solid lines). }\label{fig_b3}
\end{figure}




In this section, we present several additional numerical results. First, above studies mainly focus on the fidelity as the cost function. We now employ the energy as the cost function. The target Hamiltonian $\hat{H}^{(T)}$ is chosen as the critical 1D TFIM ($\lambda=1$) with both PBC and OBC. The system size is $N=12$. For the conventional QAOA-inspired VQE, the quantum circuit is equation~(\ref{qaoa_ising_conventional}), and the cost function is the energy $E(\textbf{x}) = \langle\psi(\textbf{x})|\hat{H}^{(T)}|\psi(\textbf{x})\rangle$. For the PRH-QAOA-inspired VQE, the quantum circuit is equation~(\ref{quantum_circuit_2}), and the cost function is the energy $E(\textbf{x},\textbf{y}) = \langle\psi(\textbf{x},\textbf{y})|\hat{H}^{(T)}|\psi(\textbf{x},\textbf{y})\rangle$ with the additional parameters $\textbf{y}$ in equations~(\ref{source_1d_ising_1}) and (\ref{source_1d_ising_2}). In figure~\ref{fig_b1}, we plot the results of $E-E_{0}$ as a function of the depth $p$, where $E$ is the parametrized $E(\textbf{x})$ or $E(\textbf{x},\textbf{y})$ obtained after optimization, and $E_{0}$ is the ground state energy of $\hat{H}^{(T)}$. The behaviors of the energy as the cost function are similar to those with the fidelity as the cost function [see figure~\ref{fig1}(a) and figure~\ref{fig2}(e)].

Second, in figures~\ref{fig1}(b) and~\ref{fig3}, for the 2D ferromagnetic TFIM with the geometry of square lattice and system size $N=3\times3$,
the strength of transverse field is chosen as $\lambda = 3.05$.  Here, we consider other values of the transverse-field
strength, including $\lambda =2.9$ and $3.2$. The results are plotted in figure~\ref{fig_b2}, and the behaviors of $1-f$ are similar to those with  $\lambda = 3.05$ [see figure~\ref{fig1}(b) and figure~\ref{fig3}(a)].

Finally, we consider another 2D ferromagnetic TFIM different from the TFIM with the geometry of square lattice. The target Hamiltonian reads $\hat{H}^{(T)} = -\sum_{\langle i,j \rangle}\hat{\sigma}_{i}^{z}\hat{\sigma}_{j}^{z} - \lambda \sum_{i=1}^{N}\hat{\sigma}_{i}^{x}$ with $\langle i,j \rangle$ referring to the nearest-neighbor sites $i$ and $j$ on the triangular lattice [see the inset of figure~\ref{fig_b3}(a)] and $\lambda=2$ being the transverse-field strength. It is noted that the Ising model with such triangular geometry has been experimentally realized in Rydberg atoms~\cite{rydberg_2d_1}. The results are plotted in figure~\ref{fig_b3}. It is seen that the PRH-QAOA can improve the performance of the VQE for 2D target Hamiltonians with triangular geometry, which may enlightens further works aiming to improve the performance of VQEs for frustrated quantum systems~\cite{2021arXiv210802175K,PhysRevB.105.094409,PhysRevB.102.075104}.


\section{Quantum circuit representation of the PRH-QAOA protocols for generating GHZ states}

\begin{figure}[]
	\centering
	\includegraphics[width=0.85\linewidth]{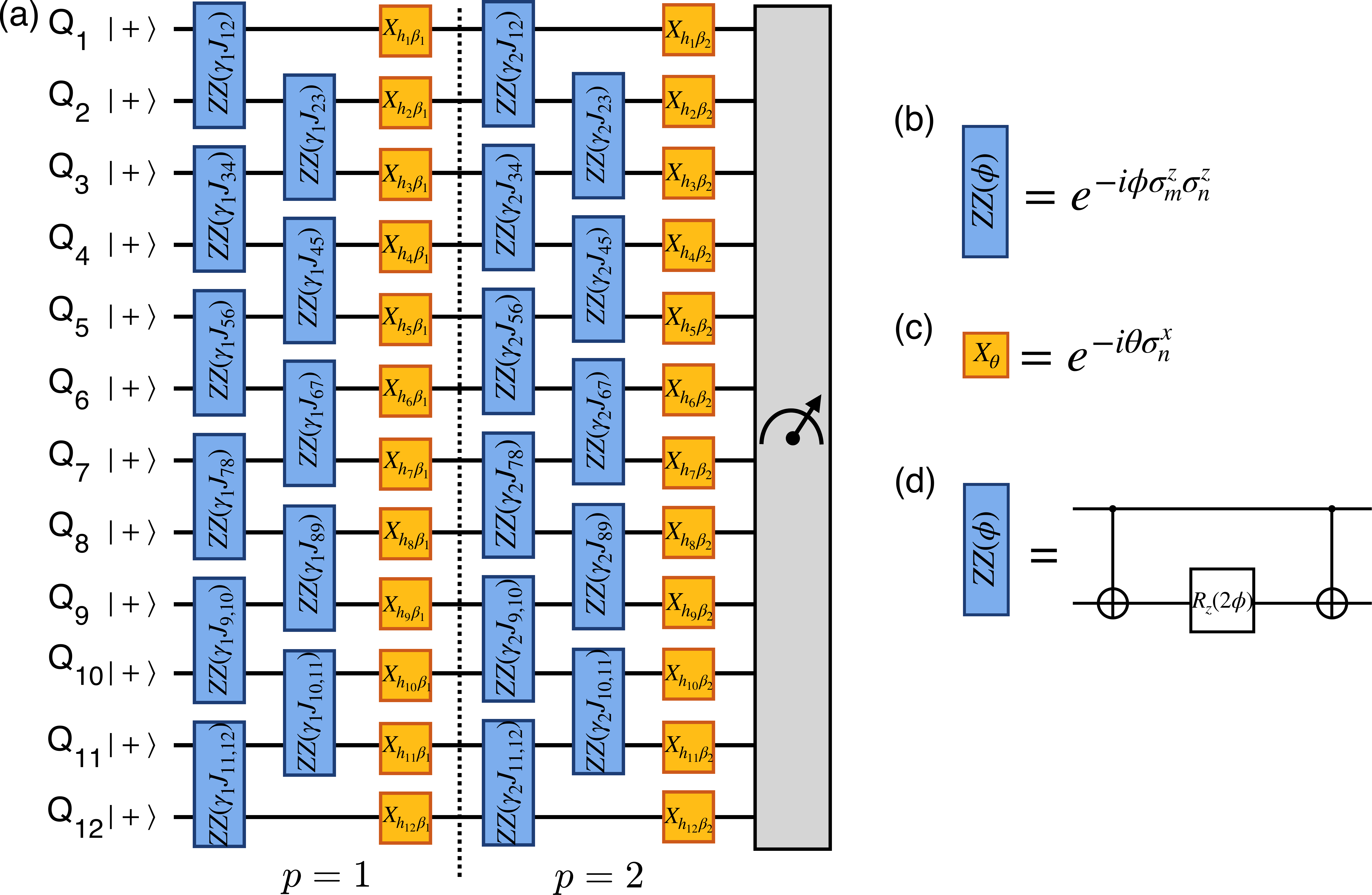}\\
	\caption{(a) The quantum circuit representation of the PRH-QAOA for generating GHZ states. Here, the 12-qubit 1D quantum circuit with OBC corresponds to the results shown in figures~\ref{fig5}(a)-(c). The initial state is $\otimes_{i=1}^{N}|+\rangle_{i}$ with $|+\rangle$ being the eigenstate of $\sigma^{x}$ with the eigenvalue $+1$. As an example, the depth of the circuit is $p=2$. The quantum circuit consists of the two-qubit gate $ZZ(\phi)$ shown in (b) and the single-qubit rotation $X_{\theta}$ shown in (c). (d) The gate $ZZ(\phi)$ can be decomposed into two CNOT gates and one single-qubit gate $R_{z}(2\phi)=e^{-i\phi\sigma^{z}}$.  }\label{fig_add1}
\end{figure}

\begin{figure}[]
	\centering
	\includegraphics[width=1\linewidth]{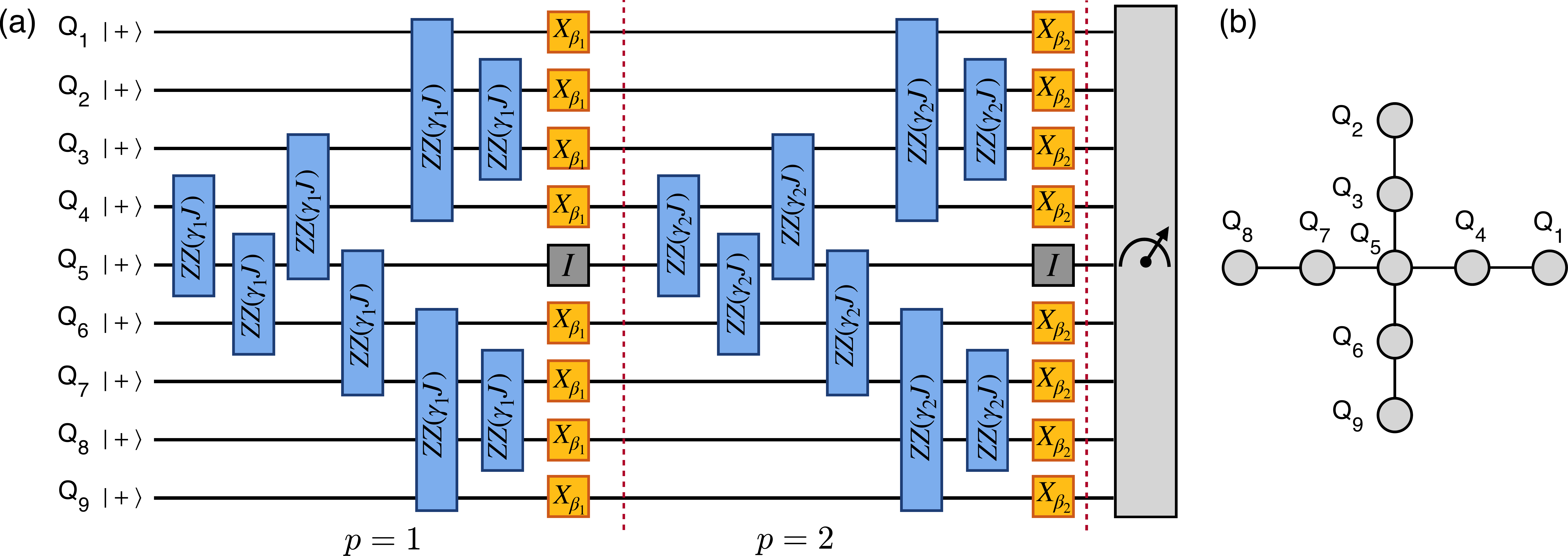}\\
	\caption{(a) The quantum circuit representation of the PRH-QAOA for generating 9-qubit GHZ states with the geometry in the figure~\ref{fig5}(f). The initial state $\otimes_{i=1}^{N}|+\rangle_{i}$, the two-qubit gate $ZZ(\phi)$, and the single-qubit rotation $X_{\theta}$ are the same as those shown in figure~\ref{fig_add1}. The gate $I$ refers to the identity gate. (b) The qubits $Q_{1}$-$Q_{9}$ are labeled, according to the results shown in the figures~\ref{fig5}(e) and (f).}\label{fig_add2}
\end{figure}

In this section, we show the quantum circuit representation of the PRH-QAOA protocols with the 1D OBC geometry (see figure~\ref{fig_add1}) and the cross-shape geometry (see figure~\ref{fig_add2}). The PRH-QAOA protocols consists of the unitary evolution under Ising Hamiltonians $U_{2}(x_{2}^{(n)}, \mathbf{J}) = \exp(ix_{2}^{(n)}\sum_{i=1}^{N}J_{i,i+1} \hat{\sigma}_{i}^{z}\hat{\sigma}_{i+1}^{z})$, i.e., the Ising dynamics, and single-qubit rotations $U_{1}(x_{1}^{(n)},\mathbf{h}) =\exp(i x_{1}^{(n)}\sum_{i=1}^{N} h_{i}\hat{\sigma}_{i}^{x})$. For digital-gate-model quantum circuit, the Ising dynamics $U_{2}(x_{2}^{(n)}, \mathbf{J})$ can be realized in two layers of two-qubit gates defined in figure~\ref{fig_add1}(b), whose decomposition is given by the quantum circuit in figure~\ref{fig_add1}(d). The single-qubit gate in figure~\ref{fig_add1}(c) directly realizes $U_{1}(x_{1}^{(n)},\mathbf{h})$.



\section{Additional numerics of generating the Greenberg-Horne-Zeilinger state with square lattice geometry}

\begin{figure}[]
	\centering
	\includegraphics[width=0.6\linewidth]{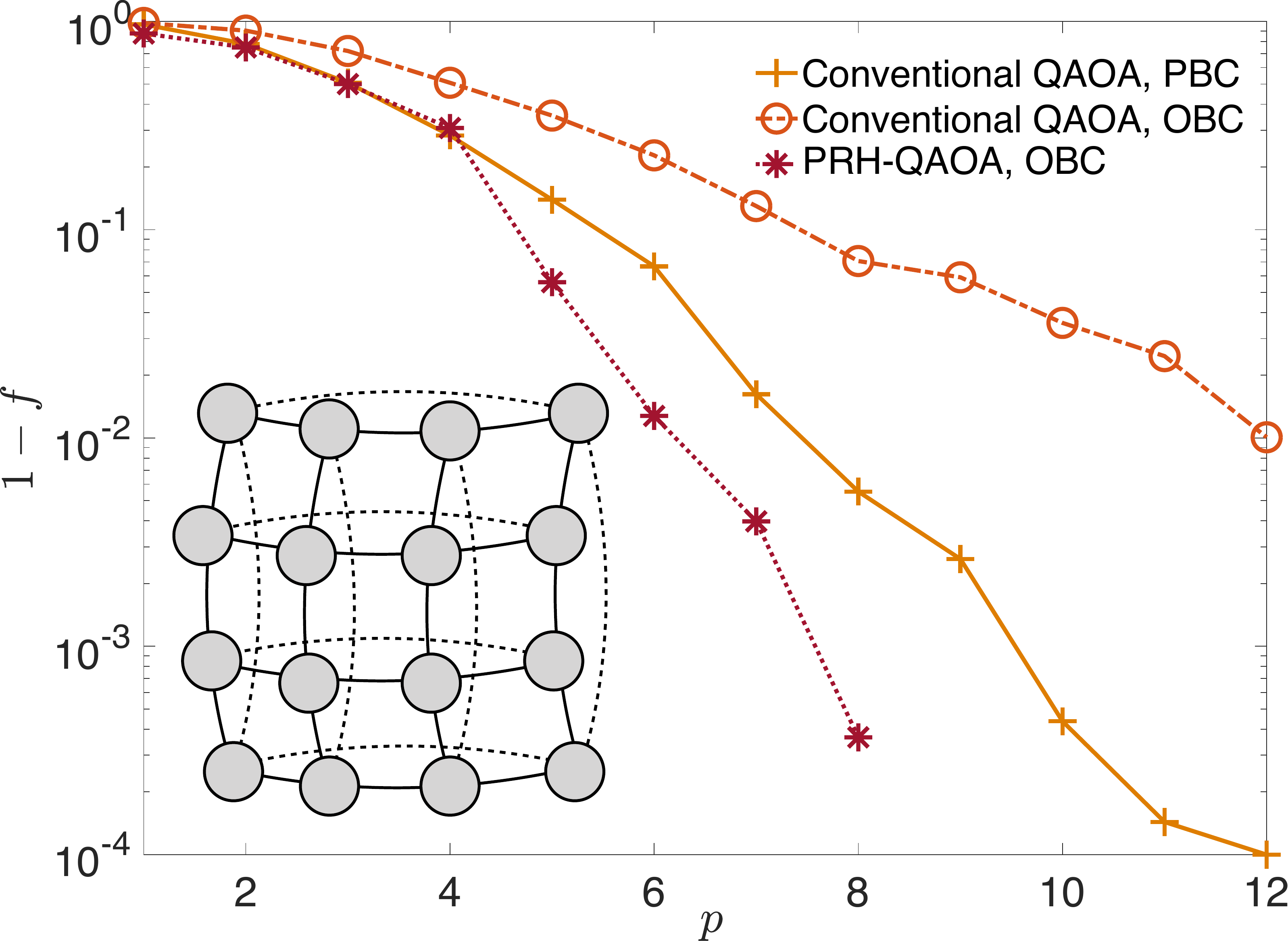}\\
	\caption{The infidelities $1-f$ with $f$ being the GHZ state fidelity, obtained after optimization, as a function of the depth $p$. The inset shows a schematic of a regular 2D lattice with $N=16$ spins (represented by circles). For the OBC case, only the interactions represented by solid lines are considered. For the PBC case, both the interactions represented by solid and dashed lines are considered. }\label{fig_c1}
\end{figure}

As shown in figure~\ref{fig7}(a), one can generate the 17-qubit GHZ state with perfect fidelity using the 4-depth quantum circuit of the QAOA with the cross-shape geometry shown in figure~\ref{fig7}(c). We now consider the Ising model on 2D square lattice with a larger system size $N=4\times 4$ as the resource Hamiltonian, and to generate 16-qubit GHZ state, both the conventional QAOA and PRH-QAOA are studied. The results are plotted in figure~\ref{fig_c1}. It is seen that for the depth of quantum circuit $p\leq 4$, all of the conventional QAOA with the both OBC and PBC quantum circuits, as well as the PRH-QAOA cannot achieve the perfect fidelity of the 16-qubit GHZ state. When $p>4$, the PRH-QAOA can generate higher fidelity GHZ states than those generated by the convention QAOA. Nevertheless, employing the Ising model on 2D square lattice as the resource Hamiltonian, for generating GHZ states, the performance of QAOA is less efficient than that employing the the cross-shape geometry shown in figures~\ref{fig7}(b) and (c).

\begin{figure}[]
	\centering
	\includegraphics[width=0.5\linewidth]{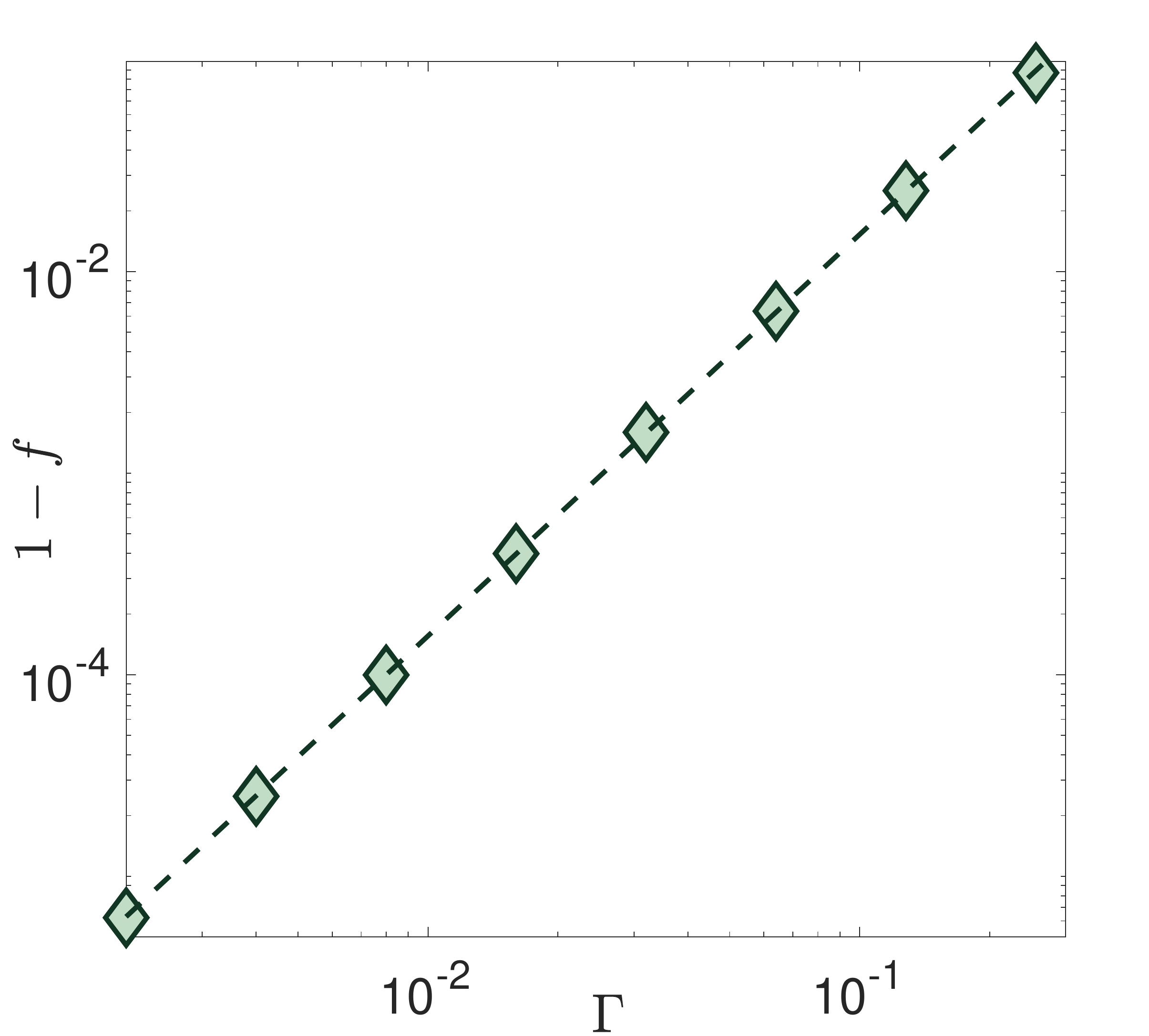}\\
	\caption{The infidelities $1-f$ as a function of the damping strength $\Gamma$. The dashed line is the linear fitting for $\log(1-f)$ and $\log\Gamma$, whose slope is $1.992$.}\label{fig_add3}
\end{figure}

\section{The effect of decoherence}
In this section, we numerically study the effect of decoherence on the PRH-QAOA for generating the GHZ state. In particular, we consider the quantum circuit with the cross-shape geometry and the system size $N=9$. As shown in figures~\ref{fig5}(d)-(f), the depth of the circuit for achieving the perfect fidelity is $p=2$. Here, we consider the damping as an example of the effect of decoherence.

To simulate the Ising dynamics with damping effect, we consider the Lindblad master equation $\frac{\text{d} \hat{\rho}(t)}{\text{d} t} = -i [\hat{H}_{zz},\hat{\rho}(t)] + \frac{1}{2}\sum_{j=1}^{N}[2\hat{C}_{j}\hat{\rho}(t)\hat{C}_{j}^{\dagger} - \{ \hat{C}_{j}^{\dagger}\hat{C}_{j}, \hat{\rho}(t) \} ]$, where $\hat{H}_{zz} = -\sum_{\langle i,j\rangle}\hat{\sigma}_{i}^{z}\hat{\sigma}_{j}^{z}$ describing ferromagnetic Ising interactions with the cross-shape geometry shown in figure~\ref{fig5}(f), and $C_{j}=\Gamma\hat{\sigma}_{j}^{-}$ is the collapse operator for the damping effect with a strength $\Gamma$.

We display the results in figure~\ref{fig_add3}. It is seen that the infidelity has a power-law increase with the damping strength $\Gamma$, i.e., $1-f \propto \Gamma^{\alpha}$ with $\alpha\simeq 1.992$. For recently developed quantum processors, the strength of damping $\Gamma$ is near $10^{-2}$~\cite{SC_2d_1,SC_2d_2,PhysRevA.97.053803}, and according to the results shown in figure~\ref{fig_add3}, the achievable infidelity of the GHZ state is $1-f \sim 10^{-4}$.

\bibliography{reference_qaoa}

\end{document}